\documentclass[fleqn,usenatbib]{mnras}

\usepackage{newtxtext,newtxmath}
\usepackage[T1]{fontenc}
\usepackage{ae,aecompl}

\usepackage{graphicx}	% Including figure files
\usepackage{amsmath}	% Advanced maths commands
\usepackage{multicol} 
\usepackage{multirow}
\usepackage{comment}
\usepackage{array}
\usepackage{soul}
\hypersetup{colorlinks=true,citecolor=blue,linkcolor=purple,filecolor=black,runcolor=black,breaklinks=true}
\usepackage{etoolbox}
\usepackage{ae,aecompl}
\usepackage[usenames]{color}
\usepackage{hyperref}
\usepackage{threeparttable}
\usepackage{mathrsfs}
\usepackage{scalerel}
\usepackage{longtable,booktabs,threeparttablex}
\usepackage{float}

\definecolor{purple}{RGB}{160,32,240}

\definecolor{red}{RGB}{225,50,50}

\definecolor{change}{RGB}{215,25,25}

\newcommand{\HST}{\emph{HST}}
\newcommand{\JWST}{\emph{JWST}}

\newcommand{\Muv}{\ensuremath{\mathrm{M}_{\mathrm{UV}}^{ }}}

\newcommand{\Lya}{\ensuremath{\mathrm{Ly}\alpha}}
\newcommand{\zLya}{\ensuremath{z_{_{\mathrm{Ly\alpha}}}}}

\newcommand{\xHI}{\ensuremath{\mathrm{x}_{\mathrm{HI}}}}

\newcommand{\fescLyA}{\ensuremath{f_{\mathrm{esc,Ly}\alpha}}}
\newcommand{\OIIIHb}{[OIII]+H\ensuremath{\beta}}

\newcommand{\Mhalo}{\ensuremath{\mathrm{M}_{\mathrm{halo}}}}

\newcommand{\Msol}{\ensuremath{\mathrm{M}_{\odot}}}
\newcommand{\Mstar}{\ensuremath{\mathrm{M}_{\ast}}}
\newcommand{\logMstar}{\ensuremath{\log\left(\mathrm{M}_{\ast}/\mathrm{M}_{\odot}\right)}}

\newcommand{\zCII}{\ensuremath{z_{_{\mathrm{[CII]}}}}}
\newcommand{\LCII}{\ensuremath{\mathrm{L}_{_{\mathrm{[CII]}}}}}

\newcommand{\DelvLya}{\ensuremath{\Delta\mathrm{v_{_{Ly\alpha}}}}}

\newcolumntype{P}[1]{>{\centering\arraybackslash}p{#1}}
\newcommand\Tstrut{\rule{0pt}{2.6ex}}         % = `top' strut
\newcommand\Bstrut{\rule[-1.2ex]{0pt}{0pt}}   % = `bottom' strut

\title[Efficient Ly$\alpha$ Transfer of UV-Bright z$\simeq$7 Galaxies]{The ALMA REBELS Survey: Efficient Ly$\alpha$ Transmission of UV-Bright z$\simeq$7 Galaxies from Large Velocity Offsets and Broad Line Widths}

\author[R. Endsley et al.]{Ryan Endsley$^{1}$\thanks{E-mail: rendsley@email.arizona.edu},
Daniel P. Stark$^{1}$,
Rychard J. Bouwens$^{2}$,
Sander Schouws$^{2}$,
Renske Smit$^{3}$,
\newauthor
Mauro Stefanon$^{2}$,
Hanae Inami$^{4}$,
Rebecca A. A. Bowler$^{5,6}$,
Pascal Oesch$^{7,8}$,
Valentino Gonzalez$^{9,10}$, 
\newauthor
Manuel Aravena$^{11}$,
Elisabete da Cunha$^{12}$,
Pratika Dayal$^{13}$,
Andrea Ferrara$^{14}$,
Luca Graziani$^{15,16}$,
\newauthor
Themiya Nanayakkara$^{17}$,
Andrea Pallottini$^{14}$,
Raffaella Schneider$^{15,18,19,20}$,
Laura Sommovigo$^{14}$,
\newauthor
Michael Topping$^{1}$,
Paul van der Werf$^{2}$,
Anne Hutter$^{13}$
\\
$^{1}$Steward Observatory, University of Arizona, 933 N Cherry Ave, Tucson, AZ 85721 USA\\
$^{2}$Leiden Observatory, Leiden University, NL-2300 RA Leiden, Netherlands\\
$^{3}$Astrophysics Research Institute, Liverpool John Moores University, 146 Brownlow Hill, Liverpool L3 5RF, United Kingdom\\
$^{4}$Hiroshima Astrophysical Science Center, Hiroshima University, 1-3-1 Kagamiyama, Higashi-Hiroshima, Hiroshima 739-8526, Japan\\
$^{5}$Astrophysics, The Denys Wilkinson Building, University of Oxford, Keble Road, Oxford, OX1 3RH, United Kingdom\\
$^{6}$Jodrell Bank Centre for Astrophysics, Department of Physics and Astronomy, School of Natural Sciences, The University of Manchester, \\
\,\, Manchester, M13 9PL, UK\\
$^{7}$Observatoire de Gen\'{e}ve, 1290 Versoix, Switzerland\\
$^{8}$Cosmic Dawn Center (DAWN), Niels Bohr Institute, University of Copenhagen, Jagtvej 128, K{\o}benhavn N, DK-2200, Denmark\\
$^{9}$Departmento de Astronomia, Universidad de Chile, Casilla 36-D, Santiago 7591245, Chile\\
$^{10}$Centro de Astrofisica y Tecnologias Afines (CATA), Camino del Observatorio 1515, Las Condes, Santiago, 7591245, Chile\\
$^{11}$Nucleo de Astronomia, Facultad de Ingenieria y Ciencias, Universidad Diego Portales, Av. Ejercito 441, Santiago, Chile\\
$^{12}$International Centre for Radio Astronomy Research, University of Western Australia, 35 Stirling Hwy., Crawley, WA 6009, Australia\\
$^{13}$Kapteyn Astronomical Institute, University of Groningen, PO Box 800, NL-9700 AV Groningen, the Netherlands\\
$^{14}$Scuola Normale Superiore, Piazza dei Cavalieri 7, 50126 Pisa, Italy\\
$^{15}$Dipartimento di Fisica, Sapienza, Universita di Roma, Piazzale Aldo Moro 5, I-00185 Roma, Italy\\
$^{16}$INAF/Osservatorio Astrofisico di Arcetri, Largo E. Femi 5, I-50125 Firenze, Italy\\
$^{17}$Centre for Astrophysics \& Supercomputing, Swinburne University of Technology, PO Box 218, Hawthorn, VIC 3112, Australia\\
$^{18}$INAF/Osservatorio Astronomico di Roma, via Frascati 33, 00078 Monte Porzio Catone, Roma, Italy\\
$^{19}$Sapienza School for Advanced Studies, Sapienza Universit\`{a} di Roma, P.le Aldo Moro 2, 00185 Roma, Italy\\
$^{20}$INFN, Sezione di Roma 1, P.le Aldo Moro 2, 00185 Roma, Italy\\
}

\date{Accepted XXX. Received YYY; in original form ZZZ}

\pubyear{2022}

\begin{document}
\label{firstpage}
\pagerange{\pageref{firstpage}--\pageref{lastpage}}
\maketitle


\begin{abstract}
Recent work has shown that UV-luminous reionization-era galaxies often exhibit strong Lyman-alpha emission despite being situated at redshifts where the IGM is thought to be substantially neutral. 
It has been argued that this enhanced Ly$\alpha$ transmission reflects the presence of massive galaxies in overdense regions which power large ionized bubbles. 
An alternative explanation is that massive galaxies shift more of their Ly$\alpha$ profile to large velocities (relative to the systemic redshift) where the IGM damping wing absorption is reduced. 
Such a mass-dependent trend is seen at lower redshifts, but whether one exists at $z\sim7$ remains unclear owing to the small number of existing systemic redshift measurements in the reionization era. 
This is now changing with the emergence of [CII]-based redshifts from ALMA. 
Here we report MMT/Binospec Ly$\alpha$ spectroscopy of eight UV-bright ($\mathrm{M_{UV}}^{}\sim-22$) galaxies at $z\simeq7$ selected from the ALMA REBELS survey. 
We detect Ly$\alpha$ in 4 of 8 galaxies and use the [CII] systemic redshifts to investigate the Ly$\alpha$ velocity profiles. 
The Ly$\alpha$ lines are significantly redshifted from systemic (average velocity offset=223 km/s) and broad (FWHM$\approx$300--650 km/s), with two sources showing emission extending to $\approx$750 km/s. 
We find that the broadest Ly$\alpha$ profiles are associated with the largest [CII] line widths, suggesting a potential link between the Ly$\alpha$ FWHM and the dynamical mass. 
Since Ly$\alpha$ photons at high velocities transmit efficiently through the $z=7$ IGM, our data suggest that velocity profiles play a significant role in boosting the Ly$\alpha$ visibility of the most UV-luminous reionization-era galaxies.
\end{abstract}


%%%%%%%%%%%%%%%%%%%%%%%%%%%%%%%%%%%%%%%%%%%%%%%%%%

%%%%%%%%%%%%%%%%% BODY OF PAPER %%%%%%%%%%%%%%%%%%

\defcitealias{Endsley2021_LyA}{E21b}

\section{Introduction} \label{sec:intro}

Reionization is a landmark event of early cosmic history, reflecting when the first luminous objects began ionizing nearly every hydrogen atom in the Universe \citep{Dayal2018,Robertson2021}.
Over the past decade, substantial progress has been made in revealing the timeline of reionization thanks to a variety of observational efforts.
The frequent detections of Ly$\alpha$ and Ly$\beta$ forests in the spectra of $z\sim6$ quasars indicate that reionization was largely complete by $z=5.9$ with an IGM neutral fraction $\xHI{} \lesssim 10$\% \citep{McGreer2015}.
Quasars at slightly higher redshifts ($z=7-7.5$) show strong Lyman-alpha damping wing features, indicating a significantly neutral IGM only $\approx$200 Myr earlier ($\xHI{} \sim 50$\%; \citealt{Mortlock2011,Greig2017,Banados2018,Davies2018,Wang2020,Yang2020_Poniuaena}).
This timeline is consistent with the reionization midpoint of $z=7.8\pm0.7$ inferred from the cosmic microwave background (CMB; \citealt{Planck2020}).

Lyman-alpha emission from high-redshift galaxies is another tool often utilized to study reionization \citep{Ouchi2020}.
Because \Lya{} resonantly interacts with HI, its observed strength is very sensitive to the ionization state of the surrounding IGM \citep[e.g.][]{MiraldaEscude1998}.
Deep spectroscopic surveys have demonstrated that the fraction of typical star-forming galaxies showing strong (rest-frame equivalent width$>$25 \AA{}) \Lya{} emission declines abruptly at $z>6$ \citep{Fontana2010,Stark2010,Ono2012,Caruana2014,Pentericci2014,Pentericci2018,Schenker2014,Jung2017,Jung2020,Hoag2019,Fuller2020} suggesting a highly neutral IGM at $z\sim7-7.5$ ($\xHI{} \gtrsim 50$\%; e.g. \citealt{Mason2018_IGMneutralFrac,Hoag2019,Jung2020,Whitler2020}).
It has also been shown that the faint end of the \Lya{} luminosity function declines faster than the UV continuum luminosity function at $z>6$ (\citealt{Hu2010,Ouchi2010,Kashikawa2011,Konno2014,Konno2018,Ota2017,Zheng2017,Itoh2018,Hu2019}; c.f. \citealt{Wold2021}), suggesting a similar reionization timeline consistent with inferences from quasars and the CMB.

Over the past few years, attention has shifted towards using \Lya{} observations to study the structure of reionization.
Recent data suggest that the \Lya{} emission strengths of UV-bright ($ \Muv \lesssim -21$) galaxies do not strongly evolve between $z\sim6$ and $z\sim7$ (\citealt{Stark2017,Endsley2021_LyA}), even though the neutrality of the IGM changes substantially over this time period.
One likely explanation for these findings is that UV-bright galaxies (which preferentially trace massive systems; e.g. \citealt{BaroneNugent2014,Harikane2018}) often reside in large ionized bubbles powered by the enhanced number of neighboring galaxies in their local overdensities (e.g. \citealt{Wyithe2005,Dayal2009,Zitrin2015,Castellano2016,Hutter2017,Weinberger2018,Endsley2021_LyA,Garaldi2021,Kannan2021,Leonova2021,Qin2021}).
The presence of these HII regions boosts \Lya{} transmission by enabling the photons to cosmologically redshift past the resonant core and into the damping wing before encountering intergalactic HI \citep{Mesinger2004,Mason2020,Smith2021}.
Reports of UV-bright \Lya{}-emitting galaxies at $z\gtrsim7$ lying in close proximity may further support the picture that large ionized structures commonly surrounded early massive galaxies \citep{Vanzella2011,Castellano2018,Jung2020,Tilvi2020,Endsley2021_LyA,Endsley2021_bubble,Hu2021}.

However, ionized bubbles are not the only mechanism capable of boosting \Lya{} transmission during reionization.
Resonant interactions within galaxies can shift \Lya{} emission redward of systemic velocity by $\gtrsim$100 km s$^{-1}$ \citep[e.g.][]{Shapley2003,Erb2014,Shibuya2014}, and thereby push the photons into the damping wing even before they escape the CGM.
If such high \Lya{} velocity offsets are common among luminous reionization-era galaxies, this would help explain their enhanced \Lya{} visibility and lessen the need for very large HII regions in their vicinity \citep[e.g.][]{Stark2017,Mason2018}.
Unfortunately, our understanding of velocity offsets at $z>6$ has remained limited by challenges in detecting not only \Lya{} from such early systems, but also a non-resonant line tracing the systemic redshift (e.g. [CII]158$\mu$m).
As a result, there are currently only four \Lya{} velocity offset measurements among extremely UV-luminous ($\Muv{} < -22$) Lyman-break selected galaxies at $z>6$ \citep{Willott2015,Stark2017,Hashimoto2019}.
Until a better census of these velocity offsets is obtained, considerable uncertainties will persist in how to connect the \Lya{} transmission of $z\gtrsim7$ galaxies to the structure of reionization.

In this work, we aim to significantly increase the number of \Lya{} velocity offset measurements among UV-luminous reionization-era galaxies.
Recently, the ongoing ALMA large program REBELS (Reionization-Era Bright Emission Line Survey; \citealt{Bouwens2021_REBELS}) yielded systemic [CII]158$\mu$m line detections from $\geq$22 UV-bright ($-23 \lesssim \Muv{} \lesssim -21.5$) galaxies at $z\geq6.5$ (Schouws et al. in prep).
Using the optical MMT/Binospec spectrograph, we have obtained \Lya{} spectra of a substantial fraction (50\%; 8/16) of [CII]-detected REBELS galaxies in the redshift range $z=6.5-7.1$.
With this combined UV+IR data set, we explore the role of velocity offsets on \Lya{} IGM transmission among UV-luminous $z\simeq7$ galaxies.
Our data additionally reveal how the broad \Lya{} line widths of these systems impact transmission through a partially neutral IGM.

This paper is organized as follows. 
In \S\ref{sec:observations}, we describe the sub-sample of $z\simeq7$ REBELS galaxies which we have thus far targeted with MMT/Binospec, as well as the details of those spectroscopic observations.
We present the \Lya{} spectra in \S\ref{sec:LyA_results}, commenting on the resulting emission line strengths and line widths.
We then utilize our ALMA [CII] detections to place the Binospec spectra into the systemic reference frame and measure \Lya{} velocity offsets (\S\ref{sec:velocity_offsets}).
In \S\ref{sec:lineProfiles}, we discuss the broad \Lya{} lines seen among our galaxies.
We then test whether our sample shows any evidence of a connection between [CII] luminosity at fixed SFR and \Lya{} equivalent width (\S\ref{sec:LCIISFR}).
Finally, we discuss how the large velocity offsets and broad line widths of massive, UV-luminous reionization-era galaxies result in efficient \Lya{} transmission through a significantly neutral IGM (\S\ref{sec:discussion}).
We summarize our main conclusions in \S\ref{sec:summary}.

In this work, we quote all magnitudes in the AB system, assume a \citet{Chabrier2003} initial mass function (IMF) with limits of 0.1--300 \Msol{}, and adopt a flat $\Lambda$CDM cosmology with parameters $h=0.7$, $\Omega_\mathrm{M}=0.3$, and $\Omega_\mathrm{\Lambda}=0.7$.

\section{Sample and Observations} \label{sec:observations}

In this work, we focus on a sample of $z\simeq7$ galaxies with recent [CII] detections from the ALMA large program REBELS \citep{Bouwens2021_REBELS,Fudamoto2021}.
REBELS (still ongoing) is targeting 40 UV-bright ($-$23 $\lesssim$ \Muv{} $\lesssim$ $-$21.5) Lyman-break selected galaxies at $z\geq6.5$.
These systems were primarily selected from the wide-area COSMOS and XMM fields ($\sim$7 deg$^2$ total). 
Here, we use the Cycle 7 data from REBELS which delivered the large majority ($\sim$85\%) of all planned observations for this program. 
The details of the ALMA data reduction and processing for [CII] and dust continuum will be described in Schouws et al. (in prep) and Inami et al. (in prep), respectively. 

We have thus far targeted \Lya{} emission in eight [CII]-detected REBELS galaxies at $z=6.5-7.1$ using the MMT/Binospec optical spectrograph \citep{Fabricant2019}.
For all observations, we utilized the 600 l/mm grating which provides sensitive coverage between $\approx$0.7--1$\mu$m at moderate resolution ($R\approx4400$).
We report the total exposure time, average seeing, wavelength coverage, and mask position angle (PA) describing the Binospec observations of each REBELS source in Table \ref{tab:binomask_info}. 
The typical exposure time per source was 5.6 hours with a typical seeing of 0.9 arcsec.
Two of our target REBELS galaxies (REBELS-14 and REBELS-39) were observed using two separate masks with slightly different wavelength coverage and PA (see Table \ref{tab:binomask_info}).
Initial \Lya{} results of REBELS-14, REBELS-15, REBELS-23, REBELS-26, and REBELS-39 were previously presented in \citet[][hereafter \citetalias{Endsley2021_LyA}]{Endsley2021_LyA} and therein identified as XMM3-227436, XMM3-504799, COS-469110, COS-534584, and COS-862541, respectively.
Here, we extend this previous work by exploring the \Lya{} emission properties of this sample in greater detail using the [CII] and dust continuum information now available from our ALMA observations.
Furthermore, we have since obtained significantly deeper (i.e. 3--5$\times$ longer exposure time) Binospec data on REBELS-14 and REBELS-39 enabling a more detailed analysis of their \Lya{} line profiles. 

Our Binospec data reduction largely follows the approach described in \citetalias{Endsley2021_LyA}, which we briefly review here.
We first process each individual exposure separately using the public Binospec data reduction pipeline \citep{Kansky2019} which performs telluric correction. 
We then coadd each exposure (for an individual mask) using the weighting scheme of \citet{Kriek2015} and apply optimal extraction \citep{Horne1986} to obtain the 1D spectra of each source.
Absolute flux calibration is determined using the spectra of multiple bright stars placed on each mask.
For REBELS-14 and REBELS-39, we coadd the data between different masks using an inverse variance weighting approach. 
Slit loss correction factors are derived by adopting the size-luminosity relation from \citet{CurtisLake2016} and assuming a S\'ersic profile with $n = 1.0$, resulting in small correction factors of $\approx$5--10\%.
We refer the interested reader to \citetalias{Endsley2021_LyA} for further details.

\begin{table}
\centering
\caption{Summary of our MMT/Binospec observations. REBELS-14 and REBELS-39 were each observed with two separate masks and we provide details for each mask.}
\begin{tabular}{P{1.6cm}P{1.0cm}P{1.0cm}P{1.6cm}P{0.8cm}} 
\hline
\multirow{2}{*}{ID} & Exposure & Seeing & Wavelength & PA \Tstrut{} \\
& Time [s] & [arcsec] & Coverage [\AA{}] & [deg] \Bstrut{} \\
\hline
REBELS-03 & 13500 & 0.73 & 7385--9909 & -37.0 \Tstrut{} \\[4pt]
REBELS-05 & 13500 & 0.73 & 6993--9516 & -37.0 \\[4pt]
\multirow{2}{*}{REBELS-14} & 22500 & 0.87 & 7548--10070 & -52.0 \\
 & 27000 & 0.91 & 7783--10306 & -12.0 \\[4pt]
REBELS-15 & 18900 & 0.91 & 7806--10330 & -116.2 \\[4pt]
REBELS-23 & 23400 & 0.99 & 7435--9959 & -98.5 \\[4pt]
REBELS-26 & 32400 & 1.09 & 7207--9731 & -98.5 \\[4pt]
REBELS-27 & 16200 & 0.94 & 7493--10017 & -98.5 \\[4pt]
\multirow{2}{*}{REBELS-39} & 7200 & 0.98 & 7250--9773 & +45.0 \\
& 14400 & 0.73 & 7355--9875 & -155.0 \\
\hline
\end{tabular}
\label{tab:binomask_info}
\end{table}

\begin{table}
\centering
\caption{Summary of the galaxy properties inferred for each of REBELS source considered in this work. Absolute UV magnitudes are reported at rest-frame 1600 \AA{}. The \OIIIHb{} EWs and stellar masses are inferred using the \textsc{beagle} SED fitting code as described in \S\ref{sec:observations}. We note that these values differ from those in the fiducial REBELS catalog (Stefanon et al. in prep) due to our alternate SED fitting approach (see text).}
\begin{tabular}{P{1.6cm}P{1.3cm}P{1.5cm}P{2.4cm}} 
\hline
ID & \Muv{} & log(\Mstar{}/\Msol{}) & \OIIIHb{} EW [\AA{}] \Tstrut{} \Bstrut{} \\
\hline
REBELS-03 & $-21.8\pm0.3$ & 8.9$^{+0.6}_{-0.6}$ & 510$^{+650}_{-320}$ \Tstrut{} \\[4pt]
REBELS-05 & $-21.6\pm0.2$ & 8.9$^{+0.7}_{-0.5}$ & 1060$^{+920}_{-530}$ \\[4pt]
REBELS-14 & $-22.7\pm0.4$ & 8.7$^{+0.4}_{-0.3}$ & 1520$^{+1300}_{-900}$ \\[4pt]
REBELS-15 & $-22.6\pm0.3$ & 9.1$^{+0.3}_{-0.2}$ & 4570$^{+1830}_{-1940}$ \\[4pt]
REBELS-23 & $-21.6\pm0.5$ & 8.8$^{+0.4}_{-0.5}$ & 830$^{+520}_{-340}$ \\[4pt]
REBELS-26 & $-21.8\pm0.1$ & 9.1$^{+0.5}_{-0.7}$ & 800$^{+640}_{-390}$ \\[4pt]
REBELS-27 & $-21.9\pm0.2$ & 9.5$^{+0.3}_{-1.0}$ & 310$^{+340}_{-210}$ \\[4pt]
REBELS-39 & $-22.7\pm0.2$ & 8.7$^{+0.3}_{-0.1}$ & 3250$^{+1010}_{-930}$ \\[4pt]
\hline
\end{tabular}
\label{tab:BEAGLEproperties}
\end{table}

The physical properties (e.g. stellar mass and \OIIIHb{} EW) of each REBELS galaxy are inferred using the SED fitting code \textsc{beagle} \citep{Chevallard2016}.
\textsc{beagle} adopts the photoionization models of star-forming galaxies from \citet{Gutkin2016} which incorporate both stellar and nebular emission by combining the latest version of the \citet{BruzualCharlot2003} stellar population synthesis models with \textsc{cloudy} \citep{Ferland2013}. 
We assume a delayed star formation history (SFR $\propto t e^{-t/\tau}$) with an allowed recent ($<$10 Myr) burst and we force star formation to have begun at least 1 Myr ago, consistent with the fitting approach of \citet{Endsley2021_OIII} and \citetalias{Endsley2021_LyA}.
While the fiducial REBELS SED fits assume a constant star formation history (Stefanon et al. in prep), our motivation for adopting this alternative `delayed+burst' fitting approach is twofold. 
First, we wish to provide a self-consistent comparison of \OIIIHb{} EWs with those inferred in \citet{Endsley2021_OIII} since it has been shown that higher \OIIIHb{} EWs connect to larger \Lya{} EWs at $z\sim7$ (\citealt{Castellano2017}; \citetalias{Endsley2021_LyA}).
Second, we aim to open up the possibility that the subset of our sources with very strong IRAC colors (and hence very large \OIIIHb{} EWs; e.g. REBELS-15 and REBELS-39) have not necessarily assembled all of their stellar mass within the past few Myr (see \citealt{Endsley2021_OIII}).
We adopt a \citet{Chabrier2003} IMF with mass limits of 0.1--300 \Msol{} and apply an SMC dust prescription \citep{Pei1992}.
During the fits, we only use photometry from bands redward of the Ly$\alpha$ break to avoid any bias due to sightline variations in IGM transmission.
Details of the near-infrared photometry used for these fits will be described in Stefanon et al. (in prep).
From our \textsc{beagle} fits, we report the median and inner 68\% confidence interval values marginalized over the output posterior probability distribution function as determined from \textsc{multinest} \citep{Feroz2008,Feroz2009}.

The [CII]-detected REBELS galaxies we have thus far followed up with Binospec are all very luminous in the rest-UV with absolute UV magnitudes of $-22.7 \leq \Muv{} \leq -21.6$ (Table \ref{tab:BEAGLEproperties}).
These values correspond to 2.5--6.9$\times$ the characteristic UV luminosity at $z\simeq7$ (M$_{\mathrm{UV}}^{\ast} = -20.6$; \citealt{Bowler2017}).
As expected from their luminous nature, our REBELS targets typically represent the most massive ($\gtrsim10^9 \Msol{}$) $z\simeq7$ galaxies known with a median inferred stellar mass of \logMstar{} = 8.9.
We note that adopting non-parametric star formation history models can increase the inferred stellar masses of our galaxies by as much as $\sim$1 dex (Topping et al. in prep), though this does not impact any of the main conclusions in this paper.
The rest-frame \OIIIHb{} equivalent widths (EWs) of our REBELS targets (inferred from the broadband SEDs) range between 310--4570 \AA{} with a median value of 940 \AA{} (see Table \ref{tab:BEAGLEproperties}).
This is similar to the typical \OIIIHb{} EW inferred for the general $z\sim7-8$ population ($\approx$760 \AA{}; \citealt{Labbe2013,Smit2014,deBarros2019,Endsley2021_OIII,Stefanon2021_colors}).

The inferred \OIIIHb{} emission strengths of our galaxies provide an estimate of their production rate of hydrogen ionizing photons, $\dot{N}_{\mathrm{ion}}$ \citep[e.g.][]{Chevallard2018_NIRSpecSimulation,Tang2019,Emami2020}, and in turn, their intrinsic \Lya{} luminosity.
With these intrinsic \Lya{} luminosities, we determine the total escape fraction of \Lya{} photons from each galaxy by comparing to the measured \Lya{} fluxes.
The intrinsic \Lya{} luminosity of each galaxy is calculated as $\mathrm{L}_{\mathrm{Ly}\alpha} = 0.677\times h\ \nu_{\mathrm{Ly}\alpha}\times \dot{N}_{\mathrm{ion}}$ where $h$ is Planck's constant and $\nu_{\mathrm{Ly}\alpha}$ is the rest-frame frequency of \Lya{}.
The factor of 0.677 is the fraction of hydrogen recombinations that result in the production of a Ly$\alpha$ photon assuming case B recombination and a temperature of 10$^4$ K \citep{OsterbrockFerland2006,Dijkstra2014}.
We note that the $\dot{N}_{\mathrm{ion}}$ values inferred from the \textsc{beagle} fits do implicitly account for dust extinction, but may be underestimated if the dust optical depth is inhomogeneous across our galaxies.
In this scenario, \OIIIHb{} emission may be much more heavily obscured from certain star-forming regions, leading to an overestimate of the \Lya{} escape fractions of our galaxies.
With our current low-resolution (beam$\approx$1.4 arcsec) ALMA data we cannot yet test for such strong spatial variations in dust attenuation within our sample, though we note that a variety of dust morphologies have been seen in UV-bright $z\sim7-8$ galaxies with higher-resolution maps \citep{Schouws2021,Bowler2022}.
We report the inferred \Lya{} escape fractions of each galaxy in \S\ref{sec:LyA_results} and estimate the role of the IGM in \S\ref{sec:discussion}.

The total star formation rates of each galaxy are determined by summing their unobscured and obscured components.
The unobscured SFRs are calculated from the UV luminosity (at 1600 \AA{} rest-frame) adopting the conversion SFR$_{\mathrm{UV}}^{}$/(\Msol{}/yr) = 7.1$\times$10$^{-29}$ L$_{\mathrm{UV}}^{}$/(erg/s/Hz) (Stefanon et al. in prep; Topping et al in prep) which results in 0.3 dex lower SFR relative to the \citet{Kennicutt1998_ARAA} conversion.
To calculate the obscured SFRs, we adopt SFR$_{\mathrm{IR}}^{}$/(\Msol{}/yr) = 1.2$\times$10$^{-10}$ L$_{\mathrm{IR}}^{}$/L$_{\odot}$ (Inami et al. in prep) which yields SFRs lower by 0.16 dex relative to \citet{Kennicutt1998_ARAA}.
Here, the total infrared luminosities, L$_{\mathrm{IR}}^{}$, are computed as 14$\nu$L$_{\nu}$ where $\nu$ is frequency of the [CII] line and L$_{\nu}$ is the dust continuum flux density determined from our ALMA observations (Sommovigo et al. in prep).
For sources undetected in far-IR continuum (REBELS-03, REBELS-15, REBELS-23, and REBELS-26), we estimate their IR luminosities by calculating an average infrared excess (IRX; L$_{\mathrm{IR}}^{}$/L$_{\mathrm{UV}}^{}$) in two UV slope bins (Topping et al. in prep).
That is, we split all dust-undetected REBELS objects by their median UV slope ($\beta = -2.04$) and stack the continuum data in each bin.
A dust continuum detection is identified in the stack of redder galaxies and we apply the average IRX to each object in this bin (REBELS-23 and REBELS-26).
The stack of bluer objects still yields a non-detection and we thus adopt upper limits on their IR luminosities from the limit on their average IRX (for REBELS-03 and REBELS-15).
The resulting UV+IR SFRs of our sample span approximately 15--80 \Msol{} yr$^{-1}$ (see Table \ref{tab:table3}).

\begin{figure*}
\includegraphics{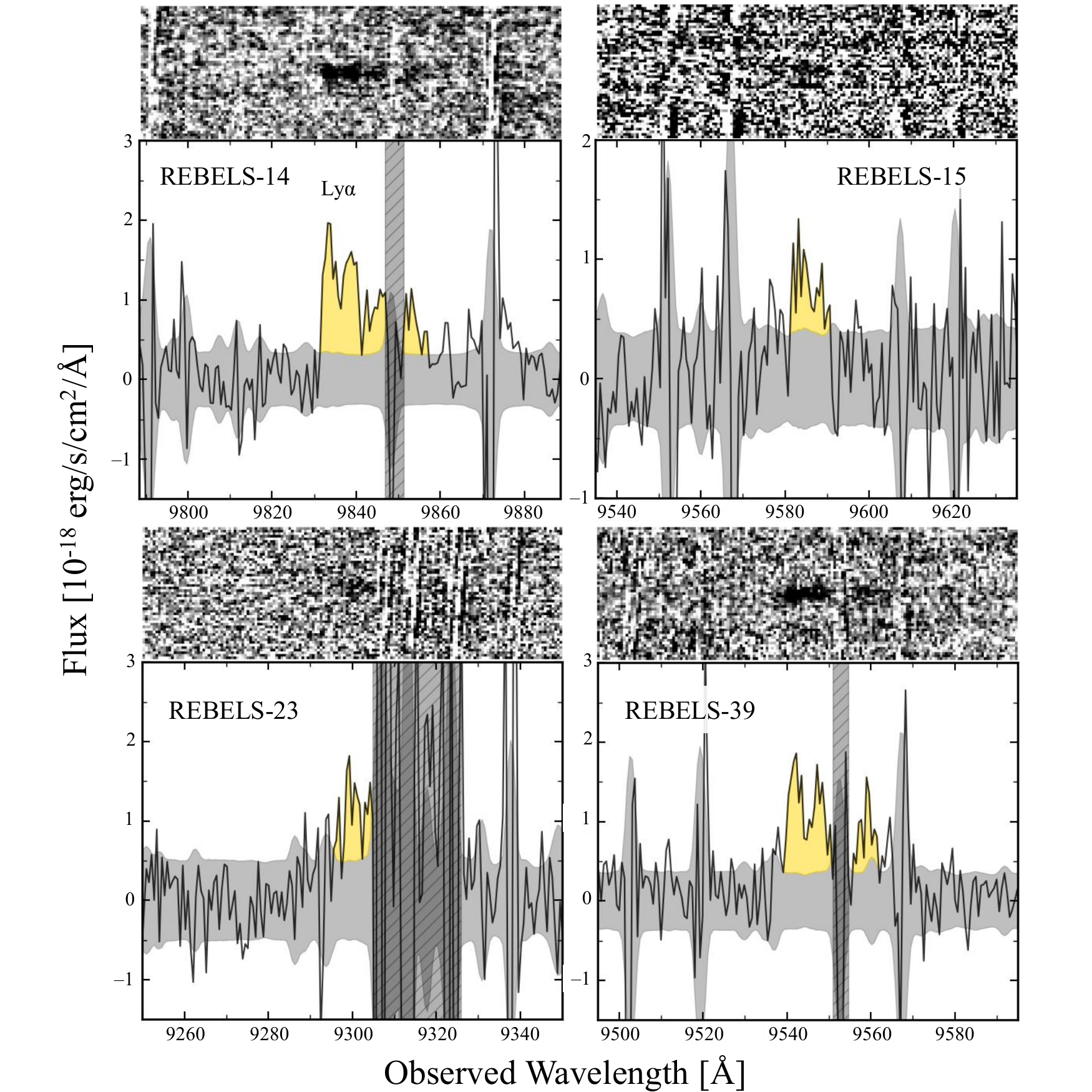}
\caption{MMT/Binospec spectra of the four REBELS galaxies with \Lya{} detections. The top panels of each sub-figure show the 2D signal-to-noise ratio maps where black is positive. The bottom panels show the 1D extraction with the 1$\sigma$ noise level in gray. For clarity, we use hashed gray regions to mask portions of the 1D \Lya{} profiles overlapping with skylines.} 
\label{fig:LyA_Detections}
\end{figure*}

\section{Lyman-alpha Spectra} \label{sec:LyA_results}

We have thus far targeted \Lya{} from eight [CII]-detected REBELS galaxies with MMT/Binospec.
In this section, we begin by detailing the Binospec spectra of the four sources in which we have confidently detected ($>$7$\sigma$) \Lya{} (\S\ref{sec:detections}) and then describe the EW limits of the remaining four sources that went undetected (\S\ref{sec:nondetections}).
For each galaxy, we calculate the \Lya{} escape fraction by comparing the observed line flux to the intrinsic \Lya{} luminosity predicted from the \textsc{beagle} SED fits as described in the previous section.
We discuss these inferred \Lya{} escape fractions in \S\ref{sec:escapeFraction}.

\subsection{Ly\boldmath{$\alpha$} Detections} \label{sec:detections}

\subsubsection{REBELS-14} \label{sec:REBELS14_LyA}

REBELS-14 is an extremely UV-luminous ($\Muv{} = -22.7$) galaxy in the XMM3 field at $\zCII{}=7.084$ (\citealt{Bouwens2021_REBELS}; Schouws et al. in prep), also referred to as XMM3-227436 in \citetalias{Endsley2021_LyA}.
This source is inferred to have a stellar mass of \logMstar{} = 8.7 and its red IRAC color ([3.6]$-$[4.5] = 0.85$^{\scaleto{+0.44}{4.5pt}}_{\scaleto{-0.37}{4.5pt}}$) suggests an \OIIIHb{} EW of 1520 \AA{}, approximately twice that typical of $z\sim7$ galaxies ($\approx$760 \AA{}; e.g. \citealt{Endsley2021_OIII}).

The Binospec \Lya{} spectrum for REBELS-14 reveals a 19.0$\sigma$ detection with a peak wavelength of 9833.3 \AA{} (Fig. \ref{fig:LyA_Detections}), corresponding to \zLya{} = 7.089 adopting a rest-frame wavelength of $\lambda_{\mathrm{Ly}\alpha} = 1215.67$ \AA{}.
We measure a total Ly$\alpha$ flux of (21.5$\pm$1.4)$\times$10$^{-18}$ erg/s/cm$^2$, corresponding to an EW of 14.6$\pm$3.0 \AA{}.
This EW is similar to the median value found among UV-bright ($-22.5 \lesssim \Muv{} \lesssim -20.5$) galaxies at $z\simeq7$ (10 \AA{}; \citetalias{Endsley2021_LyA}), indicating that the \Lya{} emission strength of REBELS-14 is fairly typical.
Throughout this work, we compute \Lya{} EWs adopting a continuum flux density from the photometric band closest to \Lya{} yet fully redward of the \Lya{} break.
For REBELS-14, this is the VIRCam \textit{J} band which yields a continuum flux density of 0.59$\pm$0.14 $\mu$Jy.
The measured \Lya{} flux from REBELS-14 suggests a \Lya{} escape fraction of $\fescLyA{}=5.0^{+2.9}_{-1.4}$\%, where we use the intrinsic line luminosity predicted from the \textsc{beagle} SED fits as described in \S\ref{sec:observations}.
We note that the escape fraction estimates in this section include the transmission of \Lya{} through the galaxy (ISM and CGM) as well as the IGM.
We also apply \Lya{} slit loss corrections assuming that the surface brightness profile tracks the rest-UV emission (see \S\ref{sec:observations}) which will underestimate the total line flux throughout the extended \Lya{} halo surrounding each galaxy.
As shown in Fig. \ref{fig:LyA_Detections}, a moderate-strength skyline overlaps with the redder portion of the \Lya{} profile from REBELS-14, possibly obscuring some of the line flux.
To estimate the potential extent of the obscuration, we assume that the line flux density in this skyline region is $1.0\times10^{-18}$ erg/s/cm$^2$/\AA{}, a value consistent with the flux density measured just outside both ends of the skyline. 
This suggests that a small fraction (14\%) of the total \Lya{} flux is obscured by this skyline which we have accounted for in the values reported above.

REBELS-14 clearly exhibits a broad asymmetric \Lya{} profile (Fig. \ref{fig:LyA_Detections}).
Given the asymmetry, we calculate the width of the line directly from the 1D spectrum, i.e. the separation between data points at half maximum flux.
To account for uncertainties, we add 100,000 realizations of noise to the 1D spectrum and take the median and 68\% confidence intervals on the derived FWHM values across all realizations.
We derive a FWHM = 640$^{+60}_{-190}$ km s$^{-1}$ where this value is corrected for the instrument resolution ($\approx$68 km s$^{-1}$).
We come back to discuss the possible physical origin of such a broad \Lya{} profile in \S\ref{sec:lineProfiles} and the implications for \Lya{} transmission during reionization in \S\ref{sec:discussion}.

\subsubsection{REBELS-15} \label{sec:REBELS15_LyA}

REBELS-15 is an extremely UV-luminous ($\Muv{} = -22.6$) galaxy in the XMM3 field at \zCII{} = 6.875 (Schouws et al. in prep) with an inferred stellar mass of \logMstar{} = 9.1.
This source was identified as XMM3-504799 in \citetalias{Endsley2021_LyA} and exhibits a very blue IRAC color ([3.6]$-$[4.5] = $-1.16^{\scaleto{+0.32}{4.5pt}}_{\scaleto{-0.43}{4.5pt}}$) suggesting an \OIIIHb{} EW = 4570 \AA{}.
At this redshift, [OIII]$\lambda$5007 only contributes slightly to the 3.6$\mu$m excess, requiring extremely strong H$\beta$ and [OIII]$\lambda$4959 emission to produce the observed IRAC color.

Our Binospec data reveal a 7.5$\sigma$ \Lya{} detection with a peak wavelength at 9583.2 \AA{} corresponding to \zLya{} = 6.883 (Fig. \ref{fig:LyA_Detections}).
The total \Lya{} flux is measured to be (5.5$\pm$1.0)$\times$10$^{-18}$ erg/s/cm$^2$ indicating an EW of 3.7$\pm$0.8 \AA{} using the VIRCam \textit{Y}-band photometry for the continuum flux density (0.58$\pm$0.09 $\mu$Jy).
REBELS-15 is thus a relatively weak Ly$\alpha$ emitter among the UV-bright $z\simeq7$ population, and is particularly weak with respect to the sub-population with strong \OIIIHb{} emission (EW$>$800 \AA{}; \citetalias{Endsley2021_LyA}).
Consistent with this result, we estimate a very low total \Lya{} escape fraction of $0.4^{+0.2}_{-0.1}$\% from REBELS-15.
The \Lya{} line profile of this source has a FWHM of 340$^{+80}_{-140}$ km s$^{-1}$.

\subsubsection{REBELS-23} \label{sec:REBELS23_LyA}

REBELS-23 is a UV-bright ($\Muv{} = -21.6$) galaxy situated in the wide-area COSMOS field at \zCII{} = 6.645 (Schouws et al. in prep) and was identified as COS-469100 in \citet{Endsley2021_OIII,Endsley2021_LyA}.
This galaxy is inferred to have a stellar mass of \logMstar{} = 8.8 and its moderately blue IRAC color ([3.6]$-$[4.5] = $-$0.53$^{\scaleto{+0.18}{4.5pt}}_{\scaleto{-0.19}{4.5pt}}$) suggests an \OIIIHb{} EW = 830 \AA{}, similar to the typical value of $z\sim7$ galaxies. 

The Binospec data reveal a 7.5$\sigma$ \Lya{} detection with a peak wavelength at 9299.3 \AA{} corresponding to \zLya{} = 6.650 (Fig. \ref{fig:LyA_Detections}). 
From the portion of the spectrum that is unobscured by strong skylines, we measure a \Lya{} flux of (9.7$\pm$1.4)$\times$10$^{-18}$ erg/s/cm$^2$ indicating an EW of 13.5$\pm$3.8 \AA{} using the VIRCam \textit{Y}-band photometry for the continuum flux density (0.27$\pm$0.07 $\mu$Jy). 
The measured line flux implies $\fescLyA{}=3.3^{+2.0}_{-1.3}$\% for REBELS-23.
Due to the patch of strong skylines redward of 9305 \AA{}, we are unable to estimate the amount of obscured line flux and thus treat the above \Lya{} flux, EW, and escape fraction measurements as lower limits.
The \Lya{} profile in the unobscured portion of the spectrum has a FWHM=330$^{+60}_{-100}$ km s$^{-1}$, though this width may also be underestimated due to the skylines.

\begin{figure}
\includegraphics{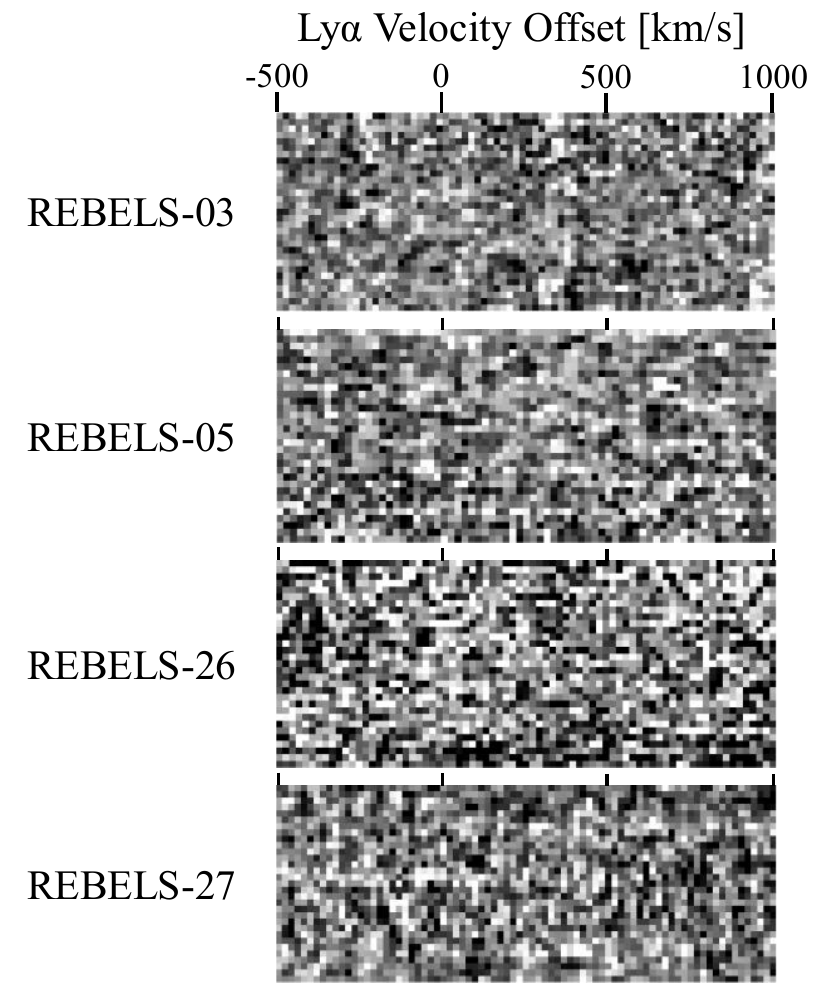}
\caption{Each panel shows the 2D signal-to-noise ratio map of the MMT/Binospec spectra of each REBELS source lacking a \Lya{} detection ($<$5$\sigma$). We show wavelengths corresponding to a conservative \Lya{} velocity offset range of \DelvLya{} = $-$500 to 1000 km s$^{-1}$ (relative to [CII]) with each panel centered on the expected spatial position of the source.}
\label{fig:LyA_nonDetections}
\end{figure}

\subsubsection{REBELS-39} \label{sec:REBELS39_LyA}

REBELS-39 is an extremely UV-luminous ($\Muv{} = -22.7$) galaxy in the COSMOS field at \zCII{} = 6.845 (\citealt{Bouwens2021_REBELS}; Schouws et al. in prep) and was identified as COS-862541 in \citet{Endsley2021_OIII,Endsley2021_LyA}.
This source has an inferred stellar mass of \logMstar{} = 8.7 and a very blue IRAC color ([3.6]$-$[4.5] = $-1.32^{\scaleto{+0.27}{4.5pt}}_{\scaleto{-0.34}{4.5pt}}$).
The IRAC color suggests an extremely large \OIIIHb{} EW of 3250 \AA{} given that this galaxy lies at a redshift where [OIII]$\lambda$5007 is only slightly transmitting through the [3.6] filter.

The Binospec data of REBELS-39 reveal a confident (15.3$\sigma$) \Lya{} detection with a peak wavelength of 9542.3 \AA{} corresponding to $\zLya{} = 6.849$ (Fig. \ref{fig:LyA_Detections}). 
We measure a total Ly$\alpha$ flux of (15.6$\pm$1.3)$\times$10$^{-18}$ erg/s/cm$^2$, indicating an EW of 10.0$\pm$1.7 \AA{} using the VIRCam \textit{Y}-band photometry for the continuum flux density (0.61$\pm$0.12 $\mu$Jy).
While this \Lya{} EW is typical of UV-bright $z\simeq7$ galaxies (\citetalias{Endsley2021_LyA}), the line flux from REBELS-39 implies a low total \Lya{} escape fraction of $1.7^{+0.4}_{-0.3}$\% given the extremely high \OIIIHb{} EW inferred for this system (3250$^{+1010}_{-930}$ \AA{}).
The \Lya{} profile for REBELS-39 does overlap with a moderate-strength skyline at $\approx$9553 \AA{}, though the Binospec data suggest that any obscuration is likely small.
Because the observed \Lya{} profile is consistent with zero flux density at both ends of the skyline (Fig. \ref{fig:LyA_Detections}), we do not introduce a correction factor for possible flux obscuration.
Similar to REBELS-14, the \Lya{} profile of REBELS-39 is extremely broad with FWHM=640$^{+60}_{-40}$ km s$^{-1}$.

\subsection{Ly\boldmath{$\alpha$} Non-detections} \label{sec:nondetections}

\subsubsection{REBELS-03} \label{sec:REBELS03_LyA}

REBELS-03 is a UV-bright ($\Muv{} = -21.8$) galaxy situated in the XMM1 field at \zCII{} = 6.969 (\citealt{Bouwens2021_REBELS}; Schouws et al. in prep).
This galaxy has an inferred stellar mass of \logMstar{} = 8.9 and a relatively weak \OIIIHb{} EW of 510 \AA{} given its flat IRAC color ([3.6]$-$[4.5] = $0.01^{\scaleto{+0.22}{4.5pt}}_{\scaleto{-0.23}{4.5pt}}$).
We have observed REBELS-03 for 3.75 hours with MMT/Binospec with clear conditions and relatively good seeing (0.73 arcsec).

We find no indication of significant ($>$5$\sigma$) \Lya{} emission from this source after searching our Binospec spectrum in conservative wavelength range corresponding to velocity offsets betweeen $-$500 to 1000 km s$^{-1}$ (Fig. \ref{fig:LyA_nonDetections}).
Upper limits on the \Lya{} EW are determined by calculating the integrated noise from the fully reduced 1D spectrum over a wavelength interval assumed relevant for the \Lya{} line.
This wavelength interval is set to begin at a value corresponding to the assumed \Lya{} redshift and spans a range equivalent to an assumed FWHM of the line.
Given the broad \Lya{} profiles observed in the four above REBELS sources, we here consider the range of FWHM=300--700 km s$^{-1}$.
For the assumed \Lya{} redshifts, we consider velocity offsets \DelvLya{} = 0--800 km s$^{-1}$ which spans the range of all robust measurements in the literature at $z>6$ (see Table \ref{tab:velocityOffsets}).
Due to the presence of a strong skylines around the expected wavelength of \Lya{}, the 5$\sigma$ EW limit for REBELS-03 is quite poor ranging from 14.7--35.5 \AA{} for the various assumed velocity offsets and FWHMs.
Here, we have used the UKIRT \textit{J}-band photometry for the continuum flux density ($m = 25.25$).
The corresponding range of 5$\sigma$ limiting \Lya{} fluxes translate to an upper limit on the total \Lya{} escape fraction of $<5.2-12.6$\%.

\begin{table*}
\centering
\caption{Summary of \Lya{} and [CII] properties of each REBELS source measured from our Binospec and ALMA observations, respectively. For sources undetected in \Lya{}, we quote the range of 5$\sigma$ EW (and associated total \Lya{} escape fraction) upper limits assuming \Lya{} redshifts corresponding to \DelvLya{} = 0--800 km s$^{-1}$ and FWHMs between 300--700 km s$^{-1}$ (see \S\ref{sec:nondetections}). For REBELS-23, we report lower limits on the \Lya{} EW, escape fraction, and FWHM given the possibility of significant skyline obscuration on the red side of the line. For sources undetected in the ALMA continuum data, we report 3$\sigma$ upper limits on their far-infrared luminosities.}
\begin{tabular}{P{1.5cm}P{0.8cm}P{0.8cm}P{1.5cm}P{1.5cm}P{1.2cm}P{1.5cm}P{1.5cm}P{1.4cm}P{1.4cm}}
\hline
\multirow{2}{*}{ID} & \multirow{2}{*}{\zCII{}} & \multirow{2}{*}{\zLya{}} & \Lya{} EW & \fescLyA{} & \DelvLya{} & \Lya{} FWHM & L$_{\mathrm{IR}}$ & \LCII{} & SFR$_{_{\mathrm{UV+IR}}}$ \Tstrut{}\Bstrut{} \\
 & & & [\AA{}] & [\%] & [km s$^{-1}$] & [km s$^{-1}$] & [10$^{11}$ L$_{\odot}$] & [10$^8$ L$_{\odot}$] & [\Msol{} yr$^{-1}$] \Tstrut{}\Bstrut{} \\
\hline
REBELS-03 & 6.969 & - & $<$14.7--35.5 & $<$5.2--12.6 & - & - & $<$2.8 & 3.2$\pm$0.6 & $16^{+7}_{-3}$ \Tstrut{} \\[4pt]
REBELS-05 & 6.496 & - & $<$3.8--4.5 & $<$0.9--1.1 & - & - & $3.3^{+2.5}_{-1.2}$ & 6.9$\pm$0.4 & $53^{+23}_{-23}$ \\[4pt]
REBELS-14 & 7.084 & 7.089 & 14.6$\pm$3.0 & $5.0^{+2.9}_{-1.4}$ & 177$^{+30}_{-30}$ & 640$^{+60}_{-190}$ & $3.4^{+2.6}_{-1.4}$ & 3.7$\pm$0.5 & $76^{+29}_{-26}$ \\[4pt]
REBELS-15 & 6.875 & 6.883 & 3.7$\pm$0.8 & $0.4^{+0.2}_{-0.1}$ & 324$^{+138}_{-34}$ & 340$^{+80}_{-140}$ & $<$3.6 & 1.9$\pm$0.3 & $34^{+16}_{-9}$ \\[4pt]
REBELS-23 & 6.645 & 6.650 & $\geq$13.5 & $\geq$3.3 & 227$^{+130}_{-92}$ & $\geq$330 & $<$3.9 & 1.3$\pm$0.2 & $24^{+12}_{-9}$ \\[4pt]
REBELS-26 & 6.598 & - & $<$5.4--6.1 & $<$1.9--2.2 & - & - & $<$4.8 & 2.0$\pm$0.4 & $28^{+5}_{-5}$ \\[4pt]
REBELS-27 & 7.090 & - & $<$5.6--13.9 & $<$3.2--8.0 & - & - & $2.9^{+2.2}_{-1.1}$ & 6.1$\pm$0.6 & $52^{+20}_{-20}$ \\[4pt]
REBELS-39 & 6.845 & 6.849 & 10.0$\pm$1.7 & $1.7^{+0.4}_{-0.3}$ & 165$^{+36}_{-35}$ & 640$^{+60}_{-40}$ & $4.3^{+3.2}_{-1.6}$ & 7.9$\pm$1.4 & $88^{+30}_{-30}$ \\[4pt]
\hline
\end{tabular}
\label{tab:table3}
\end{table*}

\subsubsection{REBELS-05} \label{sec:REBELS05_LyA}

REBELS-05 is another UV-bright ($\Muv{} = -21.6$) galaxy in the XMM1 field that lies at \zCII{} = 6.496 (\citealt{Bouwens2021_REBELS}; Schouws et al. in prep) and was first identified as ID=118717 in \citet{Bowler2014}. 
Because this source lies at $z<6.6$ where both IRAC bands are contaminated by strong nebular emission lines (i.e. \OIIIHb{} in [3.6] and H$\alpha$ in [4.5]), there is significant degeneracy between the inferred stellar mass and \OIIIHb{} EW (e.g. \citealt{Schaerer2010}).
The median posterior values from \textsc{beagle} suggest a stellar mass of \logMstar{} = 8.9 and an \OIIIHb{} EW of 1060 \AA{} given its moderately blue IRAC color ([3.6]$-$[4.5] = $-0.47^{\scaleto{+0.29}{4.5pt}}_{\scaleto{-0.35}{4.5pt}}$).
REBELS-05 was observed using the same Binospec mask as REBELS-03 with a total exposure time of 3.75 hours under clear conditions and relatively good seeing (0.73 arcsec).

We again search for \Lya{} at wavelengths corresponding to \DelvLya{} = $-$500 to 1000 km s$^{-1}$ and find no evidence of significant emission (Fig. \ref{fig:LyA_nonDetections}).
To calculate the EW upper limit, we adopt the same set of assumed velocity offsets and FWHM values as REBELS-03 and use the VIRCam \textit{Y}-band photometry for the continuum flux density ($m = 25.05$). 
This results in 5$\sigma$ limiting EWs ranging between $<$3.8--4.5 \AA{}.
Such stringent EW constraints are enabled by the fact that no strong skylines exist in the wavelength regime where \Lya{} is expected for REBELS-05.
The upper limits on \Lya{} flux translate to $\fescLyA{}<0.9-1.1$\% for REBELS-05.

\subsubsection{REBELS-26} \label{sec:REBELS26_LyA}

REBELS-26 is a UV-bright ($\Muv{} = -21.8$) galaxy in the wide-area COSMOS field at \zCII{} = 6.598 (Schouws et al. in prep) and was first identified as ID=104600 in \citet{Bowler2014}. 
Due to this redshift, the inferred stellar mass and \OIIIHb{} EW of REBELS-26 are quite degenerate similar to REBELS-05.
The median posterior values from \textsc{beagle} suggest \logMstar{} = 9.1 and \OIIIHb{} EW = 800 \AA{} for REBELS-26 given its moderate IRAC color ([3.6]$-$[4.5] = $-0.42^{\scaleto{+0.16}{4.5pt}}_{\scaleto{-0.17}{4.5pt}}$).
We have observed this galaxy for 9.0 hours with Binospec under largely clear conditions and moderate seeing on average (1.09 arcsec).

We find no indication of significant emission in our Binospec spectra for REBELS-26 at wavelengths corresponding to \DelvLya{} = $-$500 to 1000 km s$^{-1}$ (Fig. \ref{fig:LyA_nonDetections}).
Because no strong skylines exist around the expected wavelength of \Lya{}, we derive stringent 5$\sigma$ EW upper limits between 5.4--6.1 \AA{}.
Here we have used the VIRCam \textit{Y}-band photometry for the continuum flux density ($m = 24.98$). 
The total \Lya{} escape fraction of this system is inferred to be $<1.9-2.2$\% using the range of 5$\sigma$ upper limits on the line flux.

\subsubsection{REBELS-27} \label{sec:REBELS27_LyA}

REBELS-27 is another UV-luminous ($\Muv{} = -21.9$) galaxy identified across the wide-area COSMOS field located at \zCII{} = 7.090 (\citealt{Bouwens2021_REBELS}; Schouws et al. in prep).
This source was referred to as UVISTA-Y-004 in \citet{Stefanon2017_Brightestz89,Stefanon2019} and UVISTA-301 in \citealt{Bowler2020}.
It has an inferred stellar mass of \logMstar{} = 9.5 and its flat IRAC color ([3.6]$-$[4.5] = $0.10^{\scaleto{+0.20}{4.5pt}}_{\scaleto{-0.22}{4.5pt}}$) suggests relatively weak \OIIIHb{} emission (EW = 310 \AA{}).
We have observed REBELS-26 for 4.5 hours with Binospec under clear conditions and moderate seeing (0.94 arcsec).

After searching for \Lya{} at wavelengths corresponding to \DelvLya{} = $-$500 to 1000 km s$^{-1}$, we find no indication of significant line emission (Fig. \ref{fig:LyA_nonDetections}).
The 5$\sigma$ upper limiting EW ranges between 5.6--13.9 \AA{} given that a moderate-strength skyline impacts part of the relevant wavelength regime.
For the continuum flux density, we use the VIRCam \textit{J}-band photometry ($m = 24.73$).
The total \Lya{} escape fraction from REBELS-27 is inferred to be $<3.2-8.0$\%.

\subsection{Lyman-alpha Escape Fractions from REBELS Galaxies} \label{sec:escapeFraction}

It is well established that luminous, UV-selected galaxies at $z\sim2-3$ tend to have low \Lya{} escape fractions ($\approx$5\%) due to their substantial dust content and high HI covering fractions (e.g. \citealt{Hayes2010,Steidel2011,Ciardullo2014,Matthee2016,Weiss2021}).
It is, however, much less clear how efficiently \Lya{} photons are able to escape from galaxies at $z>6$.
On average, these very early systems are found to be much bluer with considerably larger sSFRs relative to typical galaxies at $z\sim2$ \citep[e.g.][]{Stark2013_NebEmission,Bouwens2014_beta,Bethermin2015,Salmon2015,Strait2020,Stefanon2021_colors}, suggesting they may have physical conditions more conducive to efficient \Lya{} escape.
Our REBELS sample enables us to investigate this possibility among the most UV-luminous ($-22.7 \leq \Muv{} \leq -21.6$) and massive ($\Mstar{} \gtrsim 10^9 \Msol{}$) galaxies at $z\sim7$.
In the previous sub-section, we quantified the \Lya{} escape fraction of each of our Binospec-targeted REBELS galaxies and the resulting values are summarized in Table \ref{tab:table3}.
Notably, the [CII] systemic redshifts enable us to place confident upper limits on the escape fraction for sources which went undetected in \Lya{} since we know whether their \Lya{} profiles may overlap with strong skylines.

None of the eight UV-luminous ($-22.7 \leq \Muv{} \leq -21.6$) $z\sim7$ REBELS galaxies which we have targeted with Binospec appear to show strong \Lya{} emission (EW$>$25 \AA{}).
This is particularly striking given that a large fraction (75\%) of these systems are inferred to exhibit high EW \OIIIHb{} emission ($\geq$800 \AA{}) implying efficient production of hydrogen ionizing photons (\citealt{Chevallard2018_z0,Tang2019}; \citetalias{Endsley2021_LyA}).
As expected from this result, we find that at least half our REBELS galaxies have low \Lya{} escape fractions of $<$2.5\% (Table \ref{tab:table3}).
Even after correcting these values for \Lya{} transmission through the IGM ($T_{\mathrm{IGM}} \approx50-80$\%; see \S\ref{sec:discussion}), we find that only 3--4\% of \Lya{} photons typically escape our REBELS galaxies, comparable to that of similarly massive ($\gtrsim 10^9 \Msol{}$) systems at $z\sim2-3$.
%Such inefficient \Lya{} escape among our REBLES sample is reasonably consistent with the theoretical expectations described in \citealt{Behrens2019}, where a simulated massive ($\Mstar{} \sim 10^{10} \Msol{}$) $z\siemq7$ galaxy was found to have an average \Lya{} escape fraction of $<$1\%.

While our Binospec-targeted REBELS galaxies are blue in the rest-UV (median $\beta = -1.94$), our ALMA data nonetheless indicate the presence of substantial dust reservoirs in many of these systems (Inami et al. in prep) which likely contribute significantly to the absorption of their \Lya{} photons (see e.g. \citealt{Behrens2019}).
It may be expected that we find a correlation between \Lya{} escape fraction and far-infrared luminosity among our REBELS sample. 
In reality, such a trend is challenging to recover with our current dataset for two reasons.
First, our ALMA data only provide shallow upper limits on the far-infrared luminosity for sources undetected in dust continuum (see Table \ref{tab:table3}), allowing for the possibility of large dust masses within these systems.
Second, our sample is limited by small statistics and a narrow dynamic range in \Lya{} escape fraction ($\lesssim$5\%).
Indeed, we find that when we separate our sample by low ($<$2.5\%) and moderate (2.5--5\%) \Lya{} escape fractions, we do not find a clear difference in the average far-infrared luminosity.
The two sources with moderate \Lya{} escape fractions (REBELS-14 and REBELS-23) have far-infrared luminosities of 3.4$\times$10$^{11}$ L$_{\odot}$ and $<$3.9$\times$10$^{11}$ L$_{\odot}$, respectively, where we quote 3$\sigma$ upper limits for non-detected objects.
Our ALMA data are consistent with similar far-infrared luminosities among the four galaxies with low ($<$2.5\%) \Lya{} escape fractions within current sensitivity limits ($\leq$3.3$\times$10$^{11}$ L$_{\odot}$ to 4.3$\times$10$^{11}$ L$_{\odot}$; see Table \ref{tab:table3}).
Here, we are ignoring the two sources with \Lya{} escape fraction upper limits $>$2.5\% (i.e. REBELS-03 and REBELS-27) since we cannot determine which bin these galaxies fall into.
Further \Lya{} observations of the REBELS sample as well as improved constraints on their far-infrared luminosity would enable a better assessment of the impact of dust on \Lya{} escape among massive ($\gtrsim 10^9$ \Msol{}) reionization-era galaxies.

\section{Analysis} \label{sec:Analysis}

All of the galaxies considered in this work have ALMA detections of their [CII]158$\mu$m emission \citep{Bouwens2021_REBELS}.
In this section, we first use the [CII] data to measure the \Lya{} velocity offsets of the four galaxies with Binospec detections, and discuss our results in context of the literature (\S\ref{sec:velocity_offsets}).
We then consider how the broad \Lya{} line widths of these four $z\simeq7$ REBELS galaxies likely assist in enhancing transmission through the IGM (\S\ref{sec:lineProfiles}).
Finally, we explore whether our sample shows any evidence of a connection between \Lya{} EW and [CII] luminosity at fixed SFR (\S\ref{sec:LCIISFR}).

\begin{figure}
\includegraphics[width=\columnwidth]{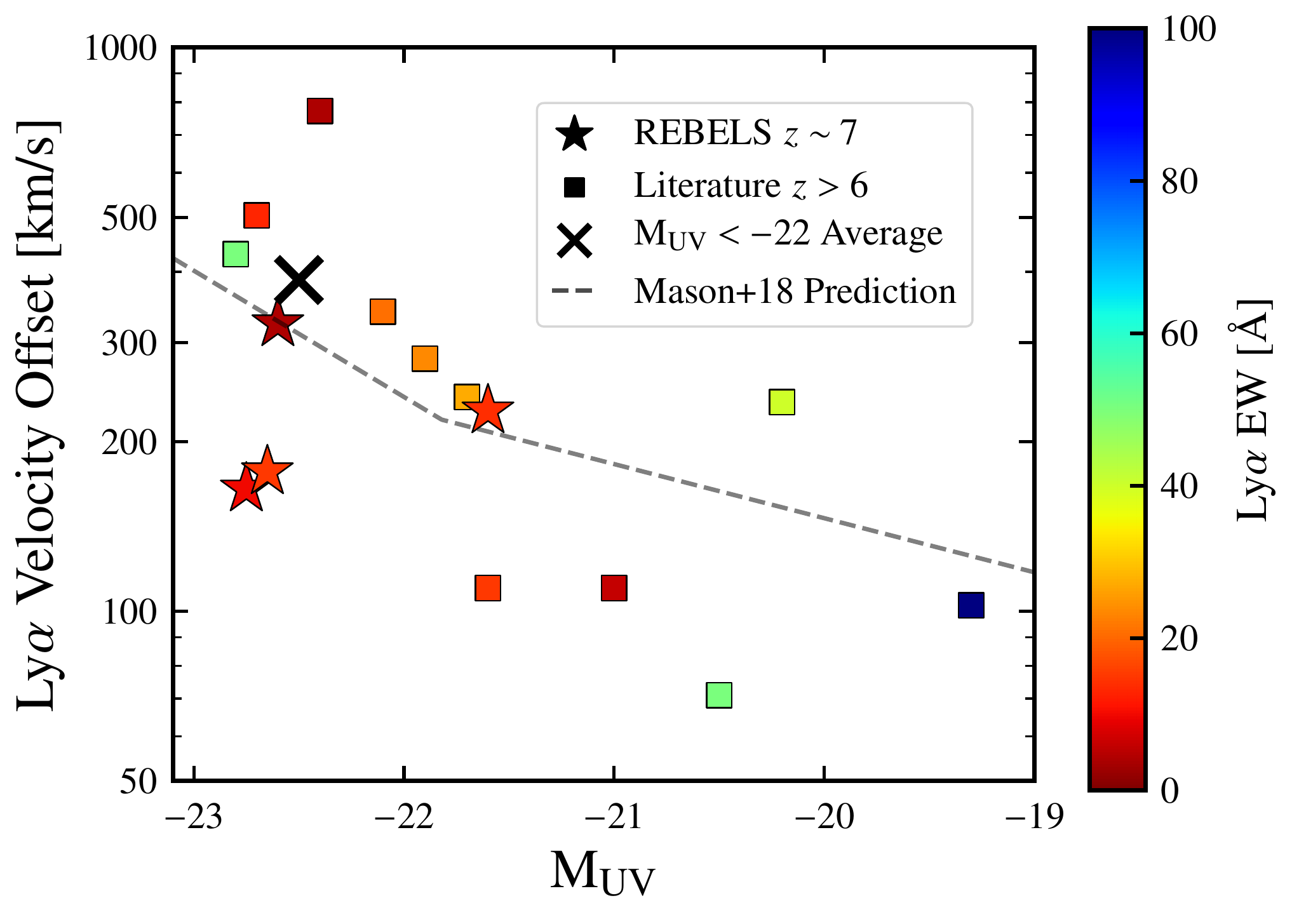}
\caption{\Lya{} velocity offsets measurements versus \Muv{} among Lyman-break selected galaxies at $z>6$. The stars show measurements from the REBELS sample while squares show additional measurements from the literature where points are color-coded by the \Lya{} EW. All values shown here are tabulated in Table \ref{tab:velocityOffsets} where, in cases when a property has multiple reported values across various works, we plot the average of those reported values. The black cross shows the average velocity offset of 387 km s$^{-1}$ measured among extremely UV-luminous ($\Muv{} < -22$) $z>6$ galaxies while the black dashed line shows the predicted relation between the average \Lya{} velocity offset and UV magnitude at $z=7$ from \citet{Mason2018_IGMneutralFrac}. Available $z>6$ velocity offset measurements at the fainter end ($\Muv{} \geq -20.5$) may be biased low given the relatively high \Lya{} EWs (40--138 \AA{}) of these three objects.}
\label{fig:vLyA_Muv}
\end{figure}

\begin{table*}
\centering
\caption{Summary of literature measurements of \Lya{} velocity offsets at z$>$6 including those from this work. We only consider measurements derived from galaxies where the \Lya{} feature is detected at S/N$>$5 and where the bluer side of the observed \Lya{} emission feature is not significantly impacted by skylines. We also ignore measurements derived from systemic line detections that were considered tentative in their published works.}
\begin{threeparttable}[t]
\begin{tabular}{P{2.0cm}P{0.7cm}P{0.7cm}P{2cm}P{1.8cm}P{7.5cm}} 
\hline
ID & $z$ & \Muv{} & \DelvLya{} & \Lya{} EW & References \Tstrut{}\Bstrut{} \\
 & & & [km s$^{-1}$] & [\AA{}] & \Bstrut{} \\
\hline
\multicolumn{6}{c}{Lyman-break Selected Galaxies} \Tstrut{} \Bstrut{} \\
CLM1 & 6.17 & $-$22.8 & 430 & 50 & \citet{Cuby2003,Willott2015} \\
WMH5 & 6.07 & $-$22.7 & 504 & 13 & \citet{Willott2013,Willott2015} \\
REBELS-15 & 6.88 & $-$22.6 & 324 & 4 & This work \\
REBELS-39 & 6.85 & $-$22.7 & 165 & 10 & This work \\
B14-65666 & 7.15 & $-$22.4 & 772 & 4 & \citet{Furusawa2016,Hashimoto2019} \\
REBELS-14 & 7.08 & $-$22.7 & 177 & 15 & This work \\
EGS-zs8-1 & 7.72 & $-$22.1 & 340 & 21 & \citet{Oesch2015,Stark2017} \\
COS-zs7-1 & 7.15 & $-$21.9 & 135--420\tnote{a} & 15--28\tnote{a} & \citet{Pentericci2016,Laporte2017,Stark2017} \\
COSMOS24108 & 6.62 & $-$21.7 & 240 & 27 & \citet{Pentericci2016,Pentericci2018} \\
REBELS-23 & 6.64 & $-$21.6 & 227 & 14 & This work \\
NTTDF6345 & 6.70 & $-$21.6 & 110 & 15 & \citet{Pentericci2011,Pentericci2016} \\
UDS16291 & 6.64 & $-$21.0 & 110 & 6 & \citet{Pentericci2016,Pentericci2018} \\
BDF-3299 & 7.11 & $-$20.5 & 71 & 50 & \citet{Vanzella2011,Maiolino2015,Carniani2017} \\
RXJ2248-ID3 & 6.11 & $-$20.2\tnote{b} & 235 & 40 & \citet{Mainali2017} \\
A383-5.2 & 6.03 & $-$19.3\tnote{b} & 85--120\tnote{c} & 138 & \citet{Stark2015_CIII,Knudsen2016} \Bstrut{} \\
\hline
\multicolumn{6}{c}{Narrowband-selected Lyman-alpha Emitting Galaxies}  \Tstrut{} \Bstrut{}  \\
VR7 & 6.53 & $-$22.4 & 213 & 38 & \citet{Matthee2019_resolvedUVandCII,Matthee2020_VR7} \Tstrut{} \\
CR7 & 6.60 & $-$22.2 & 167 & 211 & \citet{Sobral2015,Matthee2017_CR7} \\
Himiko & 6.59 & $-$21.9 & 145 & 78 & \citet{Ouchi2013,Carniani2018_Himiko} \\
\hline
\end{tabular}
\begin{tablenotes}
\item[a] Different \Lya{} redshifts and EWs are reported for COS-zs7-1 in \citet{Pentericci2016}, \citet{Laporte2017}, and \citet{Stark2017}. We list the corresponding range of velocity offsets and EWs found between the three works.
\item[b] The UV magnitudes of RXJ2248-ID3 and A383-5.2 have been corrected for gravitational lensing adopting $\mu$=5.5 and 7.3, respectively \citep{Stark2015_CIII,Mainali2017}.
\item[c] For A383-5.2, we report the \Lya{} velocity offset measured using the [CIII]$\lambda$1909 redshift from \citet{Stark2015_CIII} and the [CII]158$\mu$m redshift from \citet{Knudsen2016}. For both measurements, we adopt the \Lya{} redshift from \citet{Stark2015_CIII}.
\end{tablenotes}
\end{threeparttable}
\label{tab:velocityOffsets}
\end{table*}

\subsection{Lyman-alpha Velocity Offsets of UV-bright \boldmath{$z>6$} Galaxies} \label{sec:velocity_offsets}

An increasing number of observations have demonstrated that UV-bright ($\Muv{} \lesssim -21$) galaxies do not show strong evolution in their \Lya{} line strengths between  $z\sim6-7$ (\citealt{Ono2012,Stark2017}; \citetalias{Endsley2021_LyA}) even as the IGM neutral fraction rises rapidly over this time period (e.g. \citealt{McGreer2015,Davies2018,Wang2020,Yang2020_Poniuaena}).
This indicates that \Lya{} photons from UV-luminous $z\sim7$ galaxies are somehow able to avoid strong resonant interactions with the surrounding intergalactic HI.
One possible explanation is that their photons often escape the CGM at velocities significantly redward of systemic ($>$100 km s$^{-1}$), placing them beyond the resonant core and well into the damping wing where transmission is greatly enhanced.
While such large \Lya{} velocity offsets are commonly observed among similarly bright galaxies at $z\sim2-3$ \citep{Hashimoto2013,Erb2014,Shibuya2014}, a statistical analysis at $z>6$ has been hindered by observational challenges.
To date, only a small fraction of UV-luminous $z>6$ galaxies have been detected in \Lya{}, and much fewer have detections of a second emission line tracing the systemic redshift (e.g. [CII]158$\mu$m).
Here, we help address this issue using new [CII] detections of UV-luminous $z\geq6.5$ galaxies from the ALMA REBELS program \citep{Bouwens2021_REBELS}.

We have thus far detected \Lya{} emission in four REBELS galaxies at $\zCII{}=6.6-7.1$ (REBELS-14, REBELS-15, REBELS-23, and REBELS-39; \S\ref{sec:detections}).
All four of these systems are in the UV luminosity range where \Lya{} transmission appears to not be strongly evolving between $z\sim6-7$ ($\Muv{} \sim -22$).
We compute the \Lya{} velocity offsets of these galaxies using systemic redshifts corresponding to the central [CII] wavelength from Gaussian fits to the 1D ALMA spectra (Schouws et al. in prep).
The \Lya{} redshifts are measured from the wavelength of peak line flux (see \S\ref{sec:detections}) similar to the approach of many previous studies \citep{Maiolino2015,Stark2015_CIII,Stark2017,Carniani2017,Carniani2018_Himiko,Mainali2017,Matthee2020_VR7}.
From our redshift measurements, we calculate \Lya{} velocity offsets of 177, 324, 227, and 165 km s$^{-1}$ for REBELS-14, REBELS-15, REBELS-23, and REBELS-39, respectively (see Table \ref{tab:table3}).
Our results indicate that a significant portion of \Lya{} photons from these four UV-luminous ($\Muv{} \sim -22$) $z\sim7$ galaxies are emerging well into the damping wing (average velocity offset of 223 km s$^{-1}$) where IGM transmission is significantly boosted.

To place our results in context of previous studies, we consider the sample of $z>6$ velocity offset measurements from the literature.
Here, we only include measurements derived from secure (S/N$>$5) \Lya{} detections where the bluer side of the observed \Lya{} emission feature (containing the peak of the profile) is not significantly impacted by skylines.
We also ignore measurements derived from systemic line detections that were considered tentative in their published works.
With our REBELS sample, we have nearly doubled the number of \Lya{} velocity offset measurements among extremely UV-luminous ($\Muv{} < -22$) Lyman-break selected galaxies at $z>6$, boosting available statistics from N=4 to N=7 (see Table \ref{tab:velocityOffsets}; \citealt{Willott2015,Stark2017,Hashimoto2019}).
These seven extremely bright galaxies have an average velocity offset of 387 km s$^{-1}$.
Moreover, each exhibits \DelvLya{}$>$100 km s$^{-1}$ indicating that such large values are very common (if not ubiquitous) in the extremely UV-luminous $z>6$ population (see Fig. \ref{fig:vLyA_Muv}).

We now compare the velocity offsets seen among the extremely UV-luminous ($\Muv{} < -22$) $z>6$ galaxy population to those in the $z\sim5$ ALMA ALPINE survey \citep{LeFevre2020}.
Using the ALPINE DR1 catalog \citep{Bethermin2020,Cassata2020,Faisst2020}, we find that there are twenty $\Muv{} < -22$ galaxies with both \Lya{} and [CII] redshift measurements at $z=4.4-5.7$.
These 20 galaxies exhibit \Lya{} velocity offsets spanning $-192$ km s$^{-1}$ $\leq$ \DelvLya{} $\leq$ 520 km s$^{-1}$ with an average of 193 km s$^{-1}$.
This average value is substantially lower than that seen among the luminosity-matched $z>6$ sample (387 km s$^{-1}$).
However, we note that the velocity offsets of these ALPINE galaxies will be weighted towards low values given their relatively high \Lya{} EWs.
It has been shown that galaxies with larger \Lya{} EWs tend to exhibit smaller velocity offsets, both at $z\sim2-3$ \citep[e.g.][]{Hashimoto2013,Erb2014} as well as within the $z\sim5$ ALPINE sample \citep{Cassata2020}.
According to the ALPINE catalog, 45\% (9/20) of the extremely UV-luminous galaxies with velocity offset measurements show strong \Lya{} emission (EW$>$25 \AA{}), a considerably larger fraction than that typically reported among Lyman-break selected UV-bright ($\Muv{} < -20.5$) galaxies at $z\sim5$ ($\approx$25\%; e.g. \citealt{Stark2011,Cassata2015}).
This is to be expected given that the ALPINE galaxies were partially assembled from a sample of narrow-band selected \Lya{} emitters \citep{Faisst2020}.
We note that the sample of seven extremely UV-luminous $z>6$ galaxies considered above exhibit \Lya{} EWs which are fairly representative of this population, suggesting that their \Lya{} velocity offsets will also be representative.
Only 14\% (1/7) of these systems show strong \Lya{}, comparable to that typically seen among UV-bright ($\Muv{} < -20.5$) Lyman-break selected $z\sim6-7$ samples ($\approx$10\%; \citealt{Ono2012,Stark2011,Schenker2014,deBarros2017,Pentericci2018}).
Moreover, the median \Lya{} EW of these 7 galaxies (13 \AA{}) is very similar to that inferred for the larger UV-bright $z\sim7$ population in \citetalias{Endsley2021_LyA} ($10\pm3$ \AA{}).

Many studies have established that faint ($-20 \lesssim \Muv{} \lesssim -18$) galaxies exhibit a strong, rapid decline in \Lya{} emission strength at $z>6$ \citep[e.g.][]{Schenker2014,Pentericci2018,Fuller2020}, in contrast to the bright ($\Muv{} \lesssim -21$) population.
This luminosity dependence on \Lya{} transmission could arise (at least in part) from smaller velocity offsets among fainter $z>6$ galaxies, as may be expected if lower-mass systems have less HI gas near systemic velocity \citep[e.g.][]{Steidel2010,Erb2014}.
However, it is not clear that this is the case from existing data.
There are currently only three intrinsically faint ($\Muv{} \geq -20.5$) galaxies at $z>6$ with robust \Lya{} velocity offset measurements (see Fig. \ref{fig:vLyA_Muv} and Table \ref{tab:velocityOffsets}; \citealt{Maiolino2015,Stark2015_CIII,Carniani2017,Mainali2017}).
This limited sample size is largely the result of challenges in obtaining multi-line detections of faint reionization-era systems.
While all three galaxies do show velocity offsets smaller than the average of UV-luminous ($\Muv < -22$) systems (Table \ref{tab:velocityOffsets}), their \Lya{} emission is clearly unusual.
With EWs of 40--138 \AA{}, these galaxies fall into the rare ($\sim$5--20\%) sub-class of very strong \Lya{} emitters among the faint $z\sim6-7$ population \citep{Pentericci2018}.
Because galaxies with higher EW \Lya{} emission typically exhibit smaller velocity offsets (at least at $z<6$; e.g. \citealt{Hashimoto2013,Erb2014,Cassata2020}), the measured offsets of these three faint galaxies may be biased towards low values.
Further observations of faint $z>6$ galaxies with more typical \Lya{} emission (EW$\lesssim10$ \AA{}; \citealt{Pentericci2018}) are required to better assess the average velocity offset of this population.

\begin{figure*}
\includegraphics{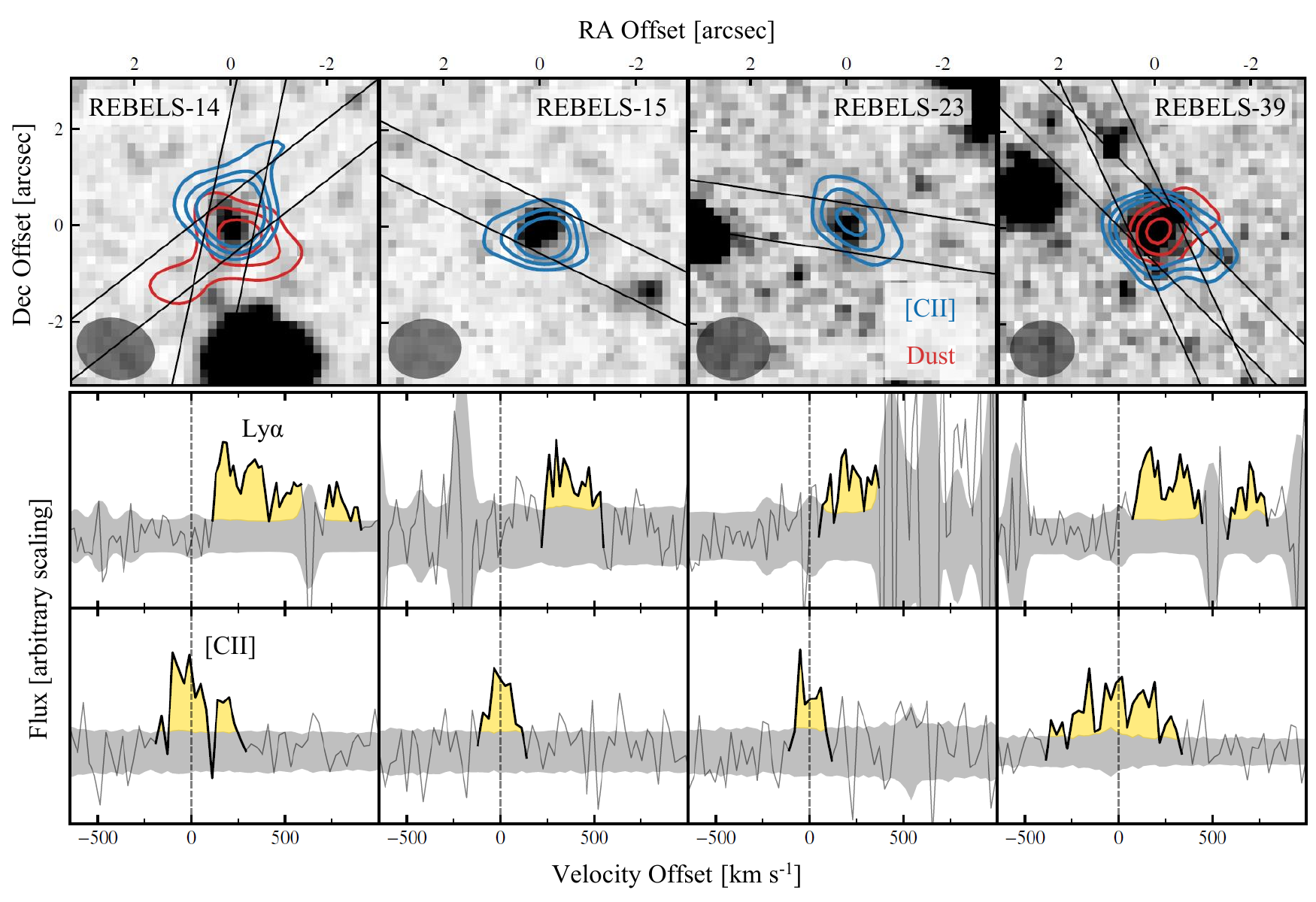}
\caption{\textbf{Top:} ALMA [CII] (blue) and dust continuum (red) contour maps overlaid on the near-infrared $\chi^2$ image of each REBELS source detected in \Lya{}. We show the 2, 3, and 4$\sigma$ contours of detections within 1 arcsec of the source centroid from near-infrared imaging. The Binospec slit positions are shown with black lines where REBELS-14 and REBELS-39 were observed with two different slit orientations. Also shown is the ALMA beam size in the lower left of each panel. \textbf{Middle:} \Lya{} velocity profiles of each source using the systemic redshift measured from [CII]. We highlight portions of the profiles free of skyline obscuration with detected flux. \textbf{Bottom:} [CII] velocity profiles of each source. Format is similar to the middle panels.}
\label{fig:ALMA_LyAOnes}
\end{figure*}

\begin{figure*}
\includegraphics{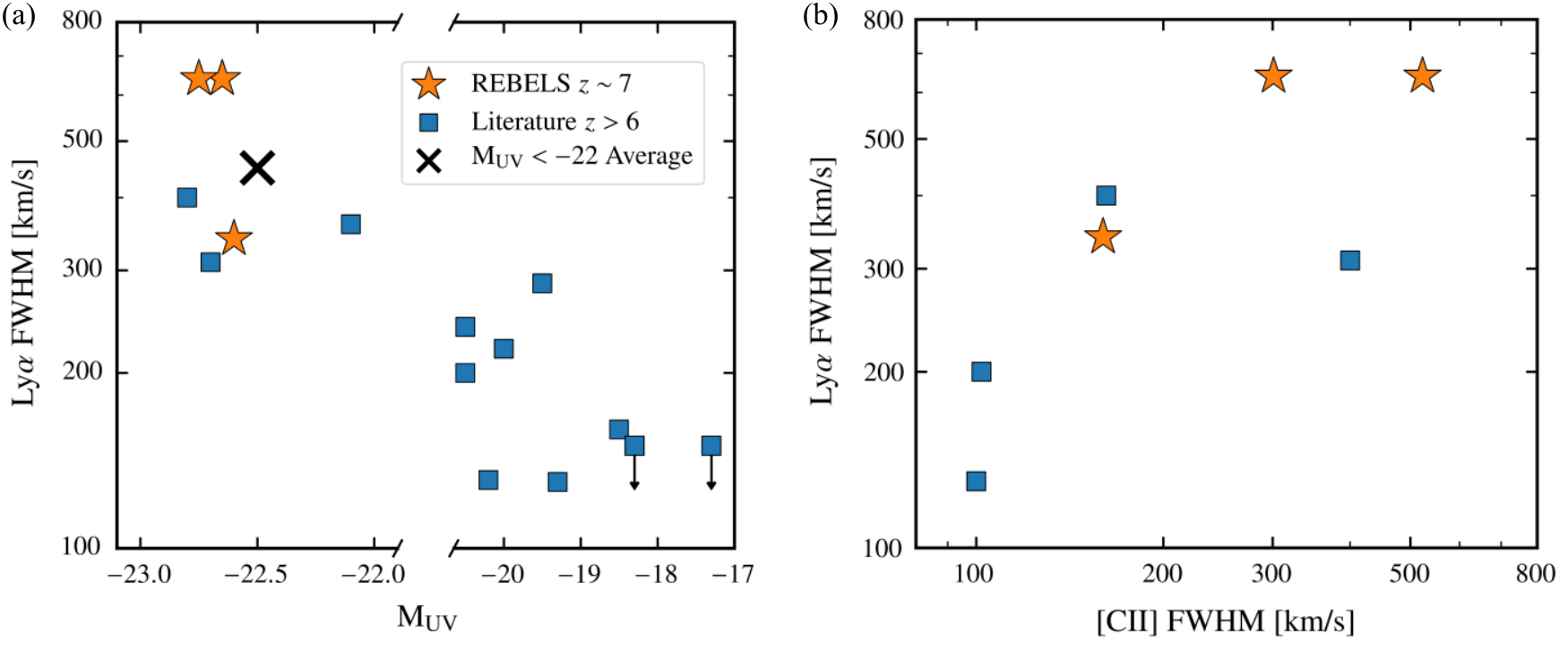}
\caption{\textbf{(a)} \Lya{} FWHM versus \Muv{} among Lyman-break selected galaxies at $z>6$. The stars represent measurements from the REBELS sample while squares show additional measurements from the literature. All values shown here are tabulated in Table \ref{tab:FWHMs} and we only consider sources with $\Muv{} < -22$ or $\Muv{} \geq -20.5$ to explore evidence of a luminosity dependence on \Lya{} FWHM at $z>6$. Current data suggest that brighter $z>6$ sources commonly exhibit larger \Lya{} FWHMs, with an average FWHM of 450 km s$^{-1}$ measured among extremely UV-luminous ($\Muv{} < -22$) galaxies (black cross). \textbf{(b)} Measurements of \Lya{} FWHM versus [CII] FWHM when available among this same sample (see Table \ref{tab:FWHMs}). These existing data suggest that the broadest \Lya{} emission is seen from galaxies with ISM reservoirs spanning the largest range of velocities.}
\label{fig:FWHM}
\end{figure*}

\begin{table*}
\centering
\caption{Literature measurements of \Lya{} FWHM among Lyman-break selected galaxies at $z>6$. We consider only those with either $\Muv{} \leq -22$ or $\Muv{} \geq -20.5$ (separated by the horizontal line) to explore a luminosity-dependent trend. This sample is limited to S/N$>$7 \Lya{} detections where the measured line width is not strongly impacted by skylines. [CII] FWHM measurements are also listed where available.}
\begin{threeparttable}[t]
\begin{tabular}{P{2.0cm}P{0.7cm}P{0.7cm}P{1.5cm}P{1.8cm}P{1.8cm}P{5.5cm}} 
\hline
ID & $z$ & \Muv{} & \Lya{} EW & \Lya{} FWHM & [CII] FWHM & References \Tstrut{} \Bstrut{} \\
 & & & [\AA{}] & [km s$^{-1}$] & [km s$^{-1}$] & \Bstrut{} \\
\hline
CLM1 & 6.17 & $-$22.8 & 50 & 400 & 162 & \citet{Cuby2003,Willott2015}  \Tstrut{} \\
WMH5 & 6.07 & $-$22.7 & 13 & 310 & 400 & \citet{Willott2013,Willott2015} \\
REBELS-15 & 6.88 & $-$22.6 & 4 & 340 & 160 & This work \\
REBELS-39 & 6.85 & $-$22.7 & 10 & 640 & 523 & This work \\
REBELS-14 & 7.08 & $-$22.7 & 15 & 640 & 301 & This work \\
EGS-zs8-1 & 7.72 & $-$22.1 & 21 & 360 & - & \citet{Oesch2015} \\ %S/N=7.2
\hline
BDF-521 & 7.01 & $-$20.5 & 64 & 240 & - & \citet{Vanzella2011} \Tstrut{} \\
BDF-3299 & 7.11 & $-$20.5 & 50 & 200 & 102 & \citet{Vanzella2011,Maiolino2015} \\
RXJ2248-ID3 & 6.11 & $-$20.2\tnote{a} & 40 & 131 & - & \citet{Mainali2017} \\
SDFJ132442 & 6.04 & $-$20.0 & 236 & 220 & - & \citet{Nagao2005} \\
RELICS-DP7 & 7.03 & $-$19.5\tnote{a} & 342 & 285 & - & \citet{Pelliccia2021} \\
A383-5.2 & 6.03 & $-$19.3\tnote{a} & 138 & 130 & 100 & \citet{Stark2015_CIII,Knudsen2016} \\
MACS0744-064 & 7.15 & $-$18.5\tnote{a} & 58 & 160 & - & \citet{Hoag2019} \\ %S/N=10
macs0717\_0859 & 6.39 & $-$18.3\tnote{a} & 45 & $\lesssim$150 & - & \citet{Vanzella2014} \\ %S/N=15
macs0717\_1730 & 6.39 & $-$17.3\tnote{a} & 32 & $\lesssim$150 & - & \citet{Vanzella2014} \\ %S/N=9
\hline
\end{tabular}
\begin{tablenotes}
\item[a] We adopt magnification factors of $\mu$=5.5, 1.15, 7.3, 3.2, 6.9, and 17.3 for RXJ2248-ID3, RELICS-DP7, A383-5.2, MACS0744-064, macs0717\_0859, and macs0717\_1730, respectively as stated by the corresponding references listed in the table.
\end{tablenotes}
\end{threeparttable}
\label{tab:FWHMs}
\end{table*}

\subsection{Lyman-alpha Line Widths of UV-luminous \boldmath{$z>6$} Galaxies} \label{sec:lineProfiles}

Along with velocity offsets, the \Lya{} line widths of $z>6$ galaxies determine how efficiently their photons transmit through a partially neutral IGM.
Galaxies with broader \Lya{} profiles emit a larger fraction of flux at high velocities where the damping wing absorption is weaker, thereby boosting IGM transmission. 
In this subsection, we discuss the \Lya{} line widths of our UV-luminous ($\Muv{} \sim -22$) REBELS galaxies and compare to those of fainter sources.

All four REBELS galaxies with Binospec detections display broad \Lya{} lines with FWHM$>$300 km s$^{-1}$ (see Table \ref{tab:table3}). 
Two of these systems (REBELS-14 and REBELS-39) show extremely wide \Lya{} profiles (FWHM$\approx$650 km s$^{-1}$) with significant emission detected $\approx$750 km s$^{-1}$ relative to systemic (see Fig. \ref{fig:ALMA_LyAOnes}). 
At these velocities, \Lya{} photons are pushed far into the damping wing where the absorption cross-section is nearly 2 orders of magnitude less than that at 100 km s$^{-1}$ \citep{Dijkstra2017}. 
The presence of Ly$\alpha$ flux at such large velocities undoubtedly enhances the visibility of these UV-luminous galaxies in a mostly neutral IGM.

We can obtain more stringent constraints on this population looking at all UV-selected $z>6$ galaxies with \Lya{} FWHM measurements.
Here, we only consider S/N$>$7 \Lya{} detections where the measured width is not strongly impacted by OH skylines.
In the sample of six UV-luminous ($\Muv{} < -22$) galaxies, we find an average \Lya{} FWHM of 450 km s$^{-1}$, with values ranging from 310-640 km s$^{-1}$ (see Fig. \ref{fig:FWHM}a and Table \ref{tab:FWHMs}, \citealt{Cuby2003,Willott2013,Oesch2015}).
If these large line widths are to preferentially boost \Lya{} transmission with respect to less luminous galaxies, we would expect to see a luminosity-dependent trend in the FWHM of \Lya{}. 
In the nine UV-selected $z>6$ galaxies with low luminosities ($\Muv{} \geq -20.5$) and published \Lya{} line width measurements, we find FWHMs ranging from 130 to 285 km s$^{-1}$ with an average of 185 km s$^{-1}$ (Table \ref{tab:FWHMs}; \citealt{Nagao2005,Vanzella2011,Vanzella2014,Stark2015_CIII,Mainali2017,Hoag2019,Pelliccia2021}). 
These are uniformly smaller than the widths of the more luminous systems considered in this paper, hinting at a luminosity-dependent trend (see Fig. \ref{fig:FWHM}a). 

The presence of such broad \Lya{} widths in the REBELS galaxies is not necessarily surprising. 
UV-luminous galaxies at $z\simeq7$ are likely to have neutral outflows with large column densities spanning a wide range of velocities, as are commonly seen in similarly luminous galaxies at lower redshifts \citep[e.g.][]{Shapley2003,Verhamme2008,Steidel2010}.
The \Lya{} profiles in these systems will accordingly take on broader widths from back-scattered emission off the far side of the outflowing gas. 
Such resonant scattering effects surely play a significant role in transferring \Lya{} photons to the large velocities described above.

Our ALMA data suggests another factor may also contribute to the broad line widths. 
The two systems in our sample exhibiting extremely broad \Lya{} emission (REBELS-14 and REBELS-39) also show atypically broad [CII] profiles with FWHM=300--520 km s$^{-1}$ (see Table \ref{tab:FWHMs}; c.f. the median FWHM=220 km s$^{-1}$ among all 24 [CII]-detected REBELS galaxies; Schouws et al. in prep), indicating that the broadest \Lya{} emission is seen from galaxies with ISM reservoirs spanning the largest range of velocities. 
This is also true in the wider literature sample (see Fig. \ref{fig:FWHM}b).
If we consider the seven UV-selected $z>6$ galaxies with \Lya{} and [CII] FWHM measurements (Table \ref{tab:FWHMs}), the three with the largest [CII] FWHM ($\geq300$ km s$^{-1}$, WMH5, REBELS-39, REBELS-14) have an average \Lya{} FWHM of 530 km/s \citep{Willott2013,Willott2015}. 
The four with lower [CII] FWHM ($\lesssim160$ km s$^{-1}$, CLM1, REBELS-15, BDF-3299, A383) have an average \Lya{} FWHM of just 270 km/s \citep{Cuby2003,Vanzella2011,Maiolino2015,Stark2015_CIII,Willott2015,Knudsen2016}.

The connection between the \Lya{} and [CII] line widths suggests that the velocity dispersion of the gas powering the line emission is likely playing a significant role in driving the FWHM of \Lya{} emission. 
The origin of the very broad [CII] emission is still not entirely clear \citep{Kohandel2019,Kohandel2020}. 
High-resolution imaging from \HST{} potentially gives some insight, revealing that UV-luminous ($\Muv{} < -22$) $z\simeq7$ galaxies tend to be composite systems comprised of several bright clumps separated by $\simeq$2--5 kpc \citep{Sobral2015,Bowler2017,Matthee2019_resolvedUVandCII}.
These clumps are likely to have significant peculiar motions ($>$100--500 km s$^{-1}$) given their separation and the dynamical masses of their host galaxies ($\simeq 1\times$10$^{10}$--3$\times$10$^{10}$ M$_\odot$), consistent with ALMA observations for a subset of these sources \citep{Jones2017,Matthee2017_CR7,Carniani2018,Hashimoto2019}.
At the spatial resolution of the REBELS [CII] maps (beam$\approx$7 kpc), these putative clumps are mostly blended, which can naturally produce the broad (and potentially multi-peaked) [CII] profiles seen in our data \citep{Kohandel2019}. 
If each clump of gas is also powering \Lya{} emission, we would expect the intrinsic \Lya{} profile to start out fairly broad in these luminous galaxies, mirroring the [CII] line profile.
Resonant scattering off of the outflowing gas will further broaden the line and shift it to higher velocities.

While fainter systems are likely to also break up into clumps, these less massive and more compact galaxies should have smaller velocity dispersions, leading to narrower initial line widths (before transfer through the outflowing gas). 
If the outflowing gas of these faint galaxies also has smaller velocities and lower column densities, it will further contribute to narrow \Lya{} widths. 
This physical picture will soon be easily testable with higher spatial resolution ALMA maps and \JWST{} IFU observations of \Lya{} and non-resonant nebular lines.
If shown to be true, it would suggest that the substantial peculiar motions of clumps in very luminous galaxies will impact the observed \Lya{} profile and in turn the atypical visibility of \Lya{} emission lines in this population. 
It would also suggest that when multiple spectral peaks in \Lya{} are observed, they do not always imply that one of the peaks is blueward of the systemic redshift (e.g. \citealt{Matthee2018,Meyer2021}). 
Such blue \Lya{} peaks are exciting as they imply a low covering fraction of neutral gas, potentially an indicator of Lyman-continuum leakage \citep[e.g.][]{Henry2015,Verhamme2015,Jaskot2019,Gazagnes2020,Hayes2021}. 
We suggest that systemic redshifts (as provided by [CII]) are critical for interpreting the \Lya{} profiles, as multiple peaks can arise naturally from blended clumps of gas in large composite systems.

\begin{figure}
\includegraphics{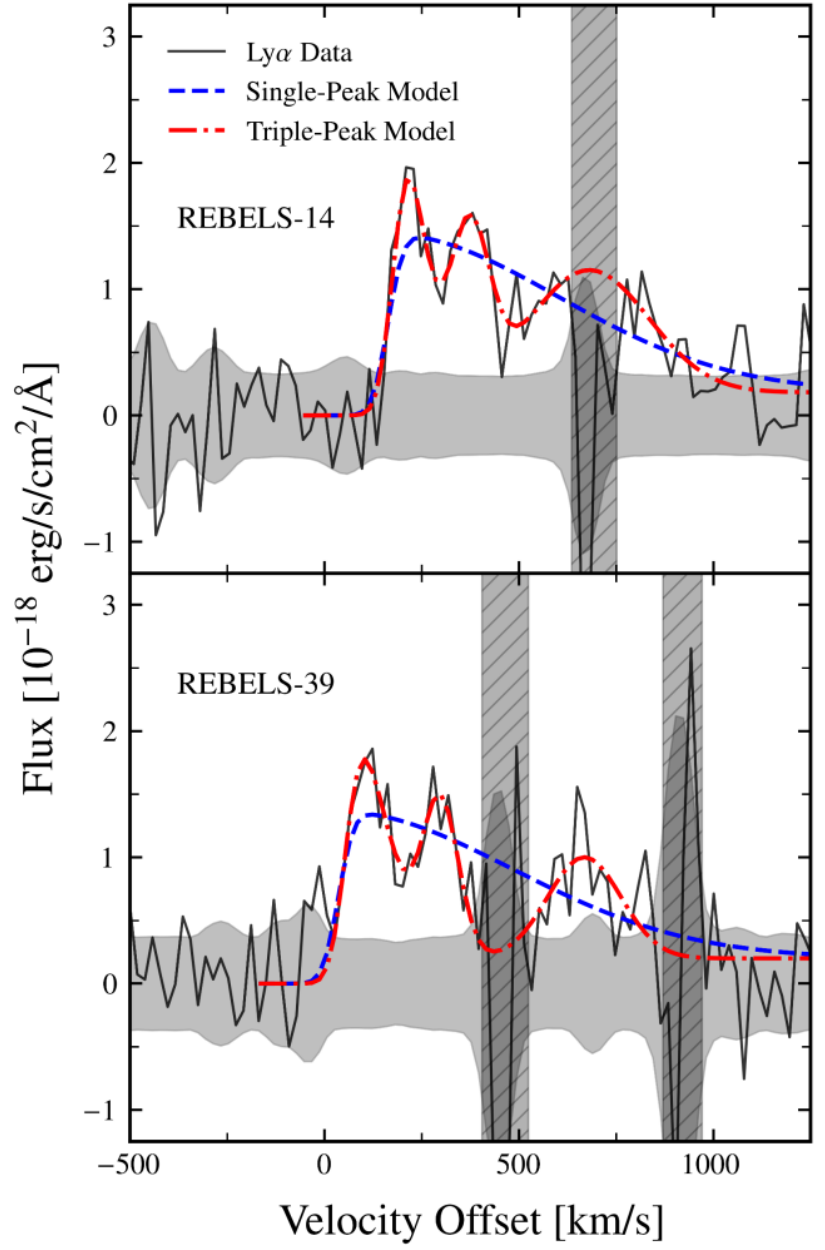}
\caption{We find tentative evidence of multi-peaked \Lya{} emission from REBELS-14 (top) and REBELS-39 (bottom). We show the MMT/Binospec data similar to Fig. \ref{fig:LyA_Detections} and overlay the best-fitting single-peak and triple-peak line profile models. Hashed gray regions show portions of the spectra overlapping with strong skylines that were not used in the fits. The triple-peak solutions provide a better match to the data (reduced $\chi^2$ = 0.50--0.66) compared to the single-peak solutions (reduced $\chi^2$ = 1.16--1.20) for both sources. These tentative multiple \Lya{} peaks (all redward of systemic) may be originating from multiple gas clumps moving with large peculiar motions within each galaxy.}
\label{fig:LyA_peakFits}
\end{figure}

Our Binospec data may be providing evidence of multi-peaked \Lya{} profiles arising from separate gas clumps in extremely luminous $z\simeq7$ galaxies.
As can be seen in Fig. \ref{fig:ALMA_LyAOnes}, the \Lya{} profiles of both REBELS-14 and REBELS-39 tentatively show multiple peaks of emission even though all flux is detected redward of the systemic velocity from [CII].
In the \Lya{} spectra of REBELS-14, there are potentially two peaks at velocities of approximately 220 and 380 km s$^{-1}$ relative to systemic where this separation of 160 km s$^{-1}$ is more than twice the resolution of our Binospec data ($\approx$68 km s$^{-1}$).
There is also a possible third \Lya{} peak in REBELS-14 giving rise to the flux seen on both sides of the moderate-strength skyline at $\approx$680 km s$^{-1}$ relative to systemic.
The \Lya{} profile of REBELS-39 shows a similar structure.
There are possibly two separate peaks at velocities of approximately 110 and 300 km s$^{-1}$ relative to systemic along with a potential third peak at $\approx$670 km s$^{-1}$.
We acknowledge that the presence of skylines and the current S/N of our data leaves significant uncertainty in the true number of \Lya{} peaks for each REBELS source, though we note that lower-redshift ($z\sim2-4$) galaxies have been shown to exhibit significant spatial variations in their \Lya{} profiles (e.g. \citealt{Erb2018,Claeyssens2019,Leclercq2020}).

To assess the plausibility of these potential multi-peak \Lya{} solutions, we fit the 1D spectra of REBELS-14 and REBELS-39 assuming three peaks of emission.
During the fits, we assume that the bluest peak is a half-Gaussian while the redder two peaks are symmetric Gaussians (due to less attenuation by the IGM/CGM at higher velocities).
Each peak component is convolved with the instrument resolution and regions overlapping with strong skylines are masked during the fits.
For comparison, we also consider single-peak profile solutions treated as a half-Gaussian convolved with the instrument resolution.
The triple-peak \Lya{} models are able to reproduce the observed line shapes for both REBELS-14 and REBELS-39, yielding best-fitting reduced $\chi^2$ values of 0.50 and 0.66 respectively.
The single-peak models provide a poorer match to the data with best-fitting reduced $\chi^2$ values of 1.16 and 1.20, respectively, where here we are accounting for the fact that the triple-peak fits have more degrees of freedom.
Nonetheless, deeper spectra will be required to verify whether multiple emission peaks (all redward of systemic velocity) are indeed present in the \Lya{} profiles of these two extremely luminous ($\Muv{} = -22.7$) $z\simeq7$ galaxies.
If shown to be true, these separate \Lya{} peaks would support the presence of multiple gas clumps with large peculiar motions within each galaxy, as may be further evidenced by their possible multi-peaked [CII] profiles (Fig. \ref{fig:ALMA_LyAOnes}).
Such a physical picture would clearly help explain the extremely broad \Lya{} emission (FWHM$\approx$640 km s$^{-1}$) observed from these two galaxies.

\subsection{The Dependence of [CII] Production on Ly\boldmath{$\alpha$} EW at \boldmath{$z\sim7$}} \label{sec:LCIISFR}

The first ALMA surveys targeting $z\gtrsim7$ galaxies resulted in very few [CII] detections \citep{Ouchi2013,Ota2014,Maiolino2015,Schaerer2015}, even though the observation depths were calibrated to local empirical relations between [CII] luminosity and SFR \citep[e.g.][]{deLooze2014}.
It has been proposed that this so-called [CII] deficit was in part due to the selection bias towards $z\gtrsim7$ galaxies with strong \Lya{} emission which had known spectroscopic redshifts (e.g. \citealt{Pentericci2016}; c.f. \citealt{Pallottini2019,Carniani2020}).
These strong \Lya{} emitters may be expected to show weaker [CII] at fixed SFR due to e.g. lower metallicities relative to the typical high-redshift galaxy population \citep[e.g.][]{Ouchi2013,Maiolino2015,Vallini2015,Pentericci2016,Ferrara2019}.  
Because our REBELS galaxies were selected via the Lyman break \citep{Bouwens2021_REBELS}, they show \Lya{} EWs more typical of bright $z\sim7$ systems and thus provide a valuable baseline for [CII] production in the reionization era.
Below, we explore whether REBELS galaxies with higher \Lya{} EWs show significantly weaker [CII] emission at fixed SFR.

We divide our REBELS galaxies into two sub-samples split by a \Lya{} EW = 10 \AA{}, which is approximately the typical EW of UV-bright $z\sim7$ galaxies (\citetalias{Endsley2021_LyA}).
For this analysis, we ignore REBELS-03 and REBELS-27 given their relatively weak upper limits on \Lya{} EW from skyline contamination.
The [CII] luminosities (Schouws et al. in prep) and UV+IR star formation rates of each galaxy (Topping et al. in prep) are reported in Table \ref{tab:table3}.

Our current sample does not show a strong trend between \Lya{} EW and [CII] luminosity at fixed SFR.
The three galaxies with relatively weak \Lya{} emission (EW$<$10 \AA{}) have [CII] luminosity to SFR ratios spanning (0.06--0.13)$\times$10$^8$ L$_{\odot}$/(M$_{\odot}$ yr$^{-1}$) with an average value of (0.09$\pm$0.04)$\times$10$^8$ L$_{\odot}$/(M$_{\odot}$ yr$^{-1}$).
These \LCII{}/SFR ratios are very similar to the three more moderate \Lya{} emitters (EW = 10--20 \AA{}) which show values spanning (0.05--0.09)$\times$10$^8$ L$_{\odot}$/(M$_{\odot}$ yr$^{-1}$) with an average of (0.06$\pm$0.02)$\times$10$^8$ L$_{\odot}$/(M$_{\odot}$ yr$^{-1}$).
Here, we are adopting a fixed conversion factor between UV luminosity and unobscured SFR as is common in the literature \citep[e.g.][]{Maiolino2015,Pentericci2016,Matthee2019_resolvedUVandCII}.
Utilizing age-dependent SFR/L$_{_{\mathrm{UV}}}$ ratios (e.g. Topping et al. in prep) does substantially increase the estimated unobscured SFRs of two sources (REBELS-15 and REBELS-39) given that their high \OIIIHb{} EWs suggest very young ages.
However, since each \Lya{} EW bin contains one of these sources, we still find that the average \LCII{}/SFR ratios are comparable for weak and moderate \Lya{} emitters in our sample.

There are two possible explanations for why we do not find significantly weaker [CII] emission (at fixed SFR) among the stronger \Lya{} emitters in our sample.
The first potential reason is that our REBELS galaxies probe a rather limited dynamic range of \Lya{} EWs ($<$20 \AA{}).
Previous studies have found evidence that relatively weak [CII] production becomes more apparent among $z\gtrsim6$ galaxies with \Lya{} EW$\gtrsim$50 \AA{} \citep[e.g.][]{Carniani2018,Harikane2018_silverrush,Matthee2019_resolvedUVandCII}.
It is possible that only at such high EWs do galaxy properties which reduce [CII] production (e.g. low metallicity or strong stellar feedback; \citealt{Ouchi2013,Vallini2015,Ferrara2019}) become significantly more common.
We also may not yet have adequate statistics to identify a modest correlation between \LCII{}/SFR ratios and \Lya{} emission at EW$\lesssim$20 \AA{}.
Further rest-UV spectroscopic follow-up of the REBELS sample would better establish if a correlation is present in this typical \Lya{} EW regime.

\section{Discussion} \label{sec:discussion}

Numerous observational campaigns have presented evidence that the IGM rapidly transitioned to a highly neutral state at $z>6$ \citep[e.g.][]{Kashikawa2011,Konno2014,McGreer2015,Greig2017,Zheng2017,Banados2018,Davies2018,Wang2020,Whitler2020,Yang2020_Poniuaena}, thereby greatly increasing the optical depth to \Lya{} given its resonant nature.
Surprisingly, UV-bright ($\Muv{} \lesssim -21$) galaxies show no evidence of strong evolution in \Lya{} emission strengths between $z\sim6$ and $z\sim7$ (\citealt{Ono2012,Stark2017}; \citetalias{Endsley2021_LyA}), raising the question of how photons from these systems were able to transmit efficiently through a partially neutral IGM.
Using our ALMA observations, we have demonstrated that UV-bright $z\sim7$ galaxies commonly exhibit both large \Lya{} velocity offsets and broad \Lya{} lines (\S\ref{sec:Analysis}).
Here, we explore how these properties enhance \Lya{} transmission in the reionization era.

\begin{figure}
\includegraphics[width=\columnwidth]{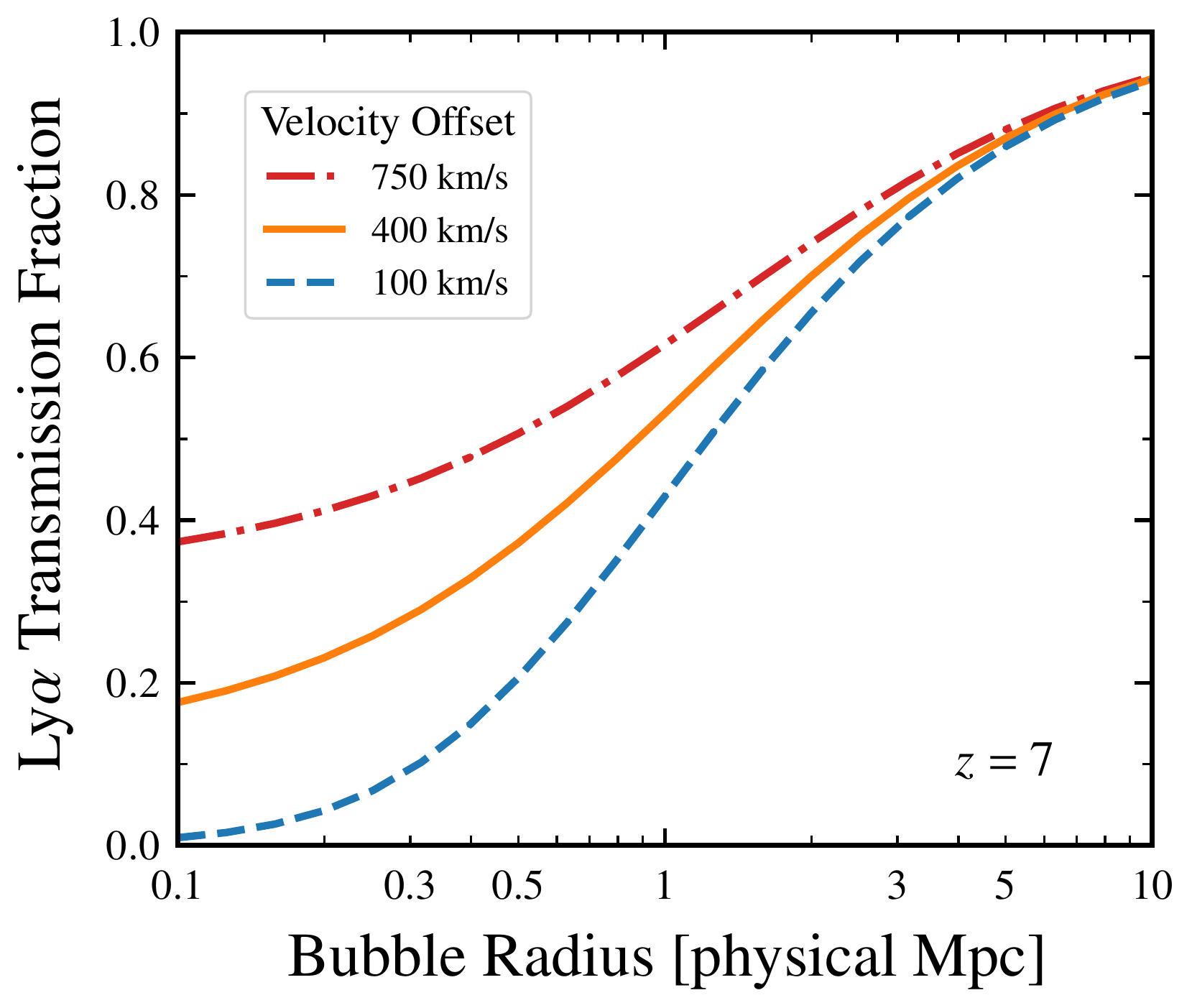}
\caption{Illustration of how \Lya{} IGM damping wing transmission during reionization depends on both the bubble size and the \Lya{} photon velocity relative to systemic. We assume that the emitting galaxy is at $z=7$ and resides in the center of a highly-ionized ($\xHI{} \lesssim 10^{-5}$) bubble with radius in physical Mpc given by the x axis. Extremely UV-luminous ($\Muv{} < -22$) galaxies at $z>6$ exhibit large velocity offsets (average $\approx$400 km s$^{-1}$) and nearly half also show very broad profiles with detectable flux extending to $\sim$750 km s$^{-1}$ (Fig. \ref{fig:ALMA_LyAOnes}). At these high velocities, \Lya{} photons transmit efficiently ($\gtrsim$40\%) through the neutral IGM even when emitted from within small ($R=0.1$ physical Mpc) bubbles.}
\label{fig:LyA_transmission}
\end{figure}

In this paper we have nearly doubled the number of \Lya{} velocity offset measurements among extremely UV-luminous ($\Muv{} < -22$) Lyman-break galaxies at $z>6$ (see Table \ref{tab:velocityOffsets}).
To quantify the expected IGM transmission of these systems, we calculate the damping wing optical depth \citep{MiraldaEscude1998} faced by \Lya{} photons at their average velocity offset ($\approx$400 km s$^{-1}$).
We assume that the IGM outside the emitting galaxy is highly ionized (\xHI{} $\lesssim$ 10$^{-5}$) out to a bubble radius $R$ beyond which the IGM is completely neutral.
In a moderate-sized bubble ($R=0.5$ physical Mpc), a typical extremely UV-luminous $z=7$ galaxy will transmit a considerable fraction ($\approx$35\%) of its \Lya{} emission at peak velocity (\DelvLya{} = 400 km s$^{-1}$, see Fig. \ref{fig:LyA_transmission}). 
Since the \Lya{} photons of these luminous systems are redshifted far past the resonant core, their IGM transmission will remain significant ($T\approx20$\%) even if situated in small ($R=0.1$ physical Mpc) bubbles.

Extremely UV-luminous ($\Muv{} < -22$) $z>6$ galaxies commonly exhibit not only large velocity offsets, but also broad \Lya{} profiles with an average FWHM = 450 km s$^{-1}$ (see Fig. \ref{fig:FWHM}a and Table \ref{tab:FWHMs}).
The combination of these two properties indicate that a substantial fraction of \Lya{} photons are frequently pushed deep into the damping wing where the HI absorption cross section is minimized.
Indeed, nearly half (3/7) of extremely UV-luminous $z>6$ galaxies with \Lya{} detections and systemic redshift measurements show \Lya{} flux extending to $\approx$750 km s$^{-1}$ (see Fig. \ref{fig:ALMA_LyAOnes} and \citealt{Hashimoto2019}), enabling \Lya{} photons transmit efficiently through the IGM even when emitted within a small ($R=0.1$ physical Mpc) ionized bubble ($T\gtrsim40\%$; see Fig. \ref{fig:LyA_transmission}).
This ability to displace \Lya{} photons to such high velocities clearly aids in the persistent visibility of massive, UV-luminous galaxies at $z>6$.

With the \Lya{} velocity profile information of our Binospec-detected REBELS galaxies, we can estimate the net IGM transmission fraction for each system assuming a given ionized bubble size.
That is, for each wavelength pixel in the observed 1D \Lya{} profiles (Fig. \ref{fig:LyA_Detections}), we divide the line flux density by the IGM transmission fraction at the corresponding velocity and re-integrate the line profile to obtain an estimate of the line flux escaping from the interstellar and circumgalactic medium.
For these UV-luminous ($-22.7 \leq \Muv{} \leq -21.6$) $z\sim7$ galaxies, we assume a fiducial bubble size of $R = 1$ physical Mpc \citep{Endsley2021_bubble} and estimate that approximately half (48--54\%) of the \Lya{} flux escaping from each galaxy has escaped through the IGM.
If these highly UV-luminous systems are instead often situated in the center of very large ($R \gtrsim 3$ physical Mpc) ionized bubbles (possibly as a result of surrounding strong galaxy overdensities; e.g. \citealt{Barkana2004,Endsley2021_bubble,Leonova2021}), we then estimate that $\gtrsim$75--80\% of their \Lya{} photons are transmitted through the IGM.
Nevertheless, we expect that the high peak \Lya{} velocity offsets and large line widths of UV-luminous $z>6$ galaxies lead to a relatively high net transmission ($\gtrsim$50\%) of \Lya{} photons through the IGM.

In contrast to bright systems, faint ($-20 \lesssim \Muv{} \lesssim -18$) galaxies show a strong decline in \Lya{} emission strengths at $z>6$ \citep{Schenker2014,Pentericci2018,Fuller2020}.
It is possible that this apparent luminosity dependence on \Lya{} transmission is (at least in part) due to smaller velocity offsets and narrower \Lya{} widths at lower UV luminosities.
While such a picture is consistent with existing observations, all these measurements come from faint $z>6$ sources with exceptionally strong \Lya{} emission (EW=32--342 \AA{}; see Tables \ref{tab:velocityOffsets} and \ref{tab:FWHMs}) thereby potentially biasing this conclusion.

Until \Lya{} detections become more accessible for typical faint reionization-era galaxies, expectations of their IGM transmission properties must be guided by data at lower redshift.
One particularly relevant result is that \Lya{} velocity offsets positively correlate with UV luminosity at $z\sim2-3$ \citep{Erb2014}.
Of course, if a similar relation holds at $z>6$, the \Lya{} emission of fainter galaxies would be more susceptible to scattering by intergalactic HI.
To better quantify the possible extent of this effect, \citet{Mason2018_IGMneutralFrac} developed a model calibrated to the $z\sim2-3$ data that predicts a galaxy's velocity offset given its UV luminosity and redshift. 
The main assumption underlying this model is that velocity offsets correlate with halo mass independently of redshift, and can therefore be linked to UV luminosity via theoretical \Muv{}--\Mhalo{} relations.
With our results, we can begin testing this model's predictions at $z\sim7$.
From the seven extremely UV-luminous ($-23 < \Muv{} < -22$) Lyman-break galaxies at $z=6-8$ with robust velocity offset measurements, we calculate an average of $\langle \DelvLya{} \rangle = 387$ km s$^{-1}$ (Table \ref{tab:velocityOffsets}).
The \citet{Mason2018_IGMneutralFrac} model predicts this empirical value within $\approx$0.1 dex ($\langle \DelvLya{} \rangle\! \approx300$ km s$^{-1}$ for $\Muv{}=-22.5$ and $z=7$), indicating reasonable agreement (see Fig. \ref{fig:vLyA_Muv}).
Assuming this model can be applied to lower luminosities, we would expect a typical faint ($\Muv{} = -19$) galaxy at $z=7$ to show a velocity offset of $\approx$100 km s$^{-1}$ which is consistent with measurements from A383-5.2 ($z=6.03$, $\Muv{} = -19.3$; \citealt{Stark2015_CIII,Knudsen2016}).
The \Lya{} IGM transmission at this velocity offset is nearly half that at $\DelvLya{} = 400$ km s$^{-1}$ assuming a moderate-sized ($R=0.5$ physical Mpc) host bubble at $z=7$ (see Fig. \ref{fig:LyA_transmission}).
The difference becomes much more substantial in smaller bubbles ($R = 0.1$ physical Mpc) where \Lya{} transmission is only $T\approx1$\% at 100 km s$^{-1}$ compared with $T\approx20$\% at 400 km s$^{-1}$.

Another relevant question is how we might expect \Lya{} FWHM to vary with UV luminosity in the reionization era.
Data at lower redshift again provides a valuable baseline.
Upon investigating a large set of literature \Lya{} observations (primarily at $z\sim0-4$), \citet{Verhamme2018} found a positive correlation between the FWHM and the velocity offset of the red \Lya{} peak.
This observed trend may arise in part from larger HI column densities both broadening the \Lya{} line as well as pushing the peak of emission to higher velocities \citep[e.g.][]{Verhamme2006,Verhamme2018,Zheng2014}.  
Outflows may also help drive this correlation since systems with gas extending to higher velocities would be expected to show both a wider \Lya{} profile and a more redshifted peak.
If fainter $z>6$ galaxies have systematically lower HI column densities or slower outflows (perhaps due to lower mass), they would thus be expected to show narrower \Lya{} lines in addition to smaller velocity offsets.
Assuming an average velocity offset of $\approx$100 km s$^{-1}$ for faint ($\Muv{} \sim -19$) galaxies at $z=7$ \citep{Mason2018_IGMneutralFrac}, the literature compilation from \citet{Verhamme2018} suggests a typical FWHM of $\approx$200-250 km s$^{-1}$.
This is consistent with measurements from RELICS-DP7, A838-5.2, and MACS0744-064 ($-19.5 \leq \Muv{} \leq -18.5$) which show an average FWHM=190 km s$^{-1}$ \citep{Stark2015_CIII,Hoag2019,Pelliccia2021}, and implies that the \Lya{} flux from faint $z=7$ galaxies would often be limited to $\lesssim$300 km s$^{-1}$ relative to systemic.
At these velocities, IGM transmission will be considerably suppressed (factor $>$3) in small ($R\sim0.1$ physical Mpc) bubbles relative to the UV-bright systems that have emission extending to $\approx$750 km s$^{-1}$.

These predictions help explain why it has been so challenging to detect \Lya{} emission from faint $z\gtrsim7$ galaxies, even those which appear to sit in the vicinity of an ionized bubble.
One likely ionized region is in the BDF field where three relatively bright ($\Muv{} \leq -20.5$) galaxies at $z=7.0-7.1$ exhibit strong \Lya{} emission \citep{Vanzella2011,Castellano2018}.
However, deep follow-up observations yielded no \Lya{} detections among all twelve fainter ($\Muv{} > -20.25$) $z\sim7$ candidates targeted in this field \citep{Castellano2018}.
While it is currently unclear whether these faint systems indeed reside in the same ionized structure(s), it is nonetheless expected that their \Lya{} photons would experience weaker transmission due to smaller velocity offsets and narrower line widths.
Using the predictions described above, an intrinsically faint ($\Muv{} \sim -19$) $z=7$ galaxy would likely have $\approx$2$\times$ lower \Lya{} transmission relative to an extremely luminous ($\Muv{} < -22$) system assuming a moderate-sized ($R=0.5$ physical Mpc) host bubble.
These expectations for reduced transmission also help explain the dearth of \Lya{} detections among faint yet lensed $z\sim7-8$ galaxies \citep{Hoag2019,Mason2019a}.

\section{Summary} \label{sec:summary}
We present MMT/Binospec \Lya{} spectra of eight UV-luminous ($-22.7 \leq \Muv{} \leq -21.6$), massive ($8.7 \leq \logMstar{} \leq 9.5$) [CII]-detected galaxies at $z=6.5-7.1$ selected from the recent ALMA Large program REBELS \citep{Bouwens2021_REBELS}.
With this data, we report four new measurements of \Lya{} velocity offsets among UV-bright reionization-era galaxies, improving our understanding of how their photons transmit efficiently through a partially neutral IGM.
We also investigate the role played by the broad \Lya{} profiles of UV-luminous $z>6$ galaxies.
Our conclusions are as follows:

\begin{enumerate}
    \item We confidently ($>$7$\sigma$) detect \Lya{} in four of eight [CII]-detected REBELS galaxies that we have so far targeted with MMT/Binospec. The \Lya{} EWs of our detected sources range between 3.7--14.6 \AA{} which are all typical of massive UV-luminous $z\simeq7$ galaxies (\citetalias{Endsley2021_LyA}). For two of the sources lacking a \Lya{} detection, we place stringent 5$\sigma$ EW limits of $\leq$6 \AA{} given the lack of skylines in the relevant part of the spectrum.
    
    \item Among the four REBELS galaxies with Binospec detections, we measure \Lya{} velocity offsets of 165, 177, 227, and 324 km s$^{-1}$. Our sample nearly doubles the number of velocity offset measurements among extremely UV-luminous ($\Muv{} < -22$) Lyman-break selected galaxies at $z>6$. All seven of these systems show large ($>$100 km s$^{-1}$) velocity offsets, indicating that their \Lya{} photons are pushed well past the strong resonant core of the HI absorption cross section. At their average velocity offset of 387 km s$^{-1}$, \Lya{} photons transmit efficiently through the IGM with $T\approx35$\% when emerging from a $z=7$ galaxy situated in a moderate-sized ($R = 0.5$ physical Mpc), highly-ionized ($\xHI{} < 10^{-5}$) bubble.
    
    \item All four REBELS galaxies with Binospec detections display broad \Lya{} lines with FWHM$>$300 km s$^{-1}$, indicating that a large fraction of their photons are at high velocities where damping wing absorption is weaker. Two of these galaxies show extremely broad \Lya{} profiles (FWHM=640 km s$^{-1}$) with significant emission extending to $\approx$750 km s$^{-1}$ relative to systemic. At such high velocities, \Lya{} photons transmit efficiently ($T\gtrsim40$\%) through the IGM, even when the emitting source resides in a small ($R=0.1$ physical Mpc) ionized bubble. Broad \Lya{} profiles (FWHM=300-400 km s$^{-1}$) are also observed from the three other UV-luminous ($\Muv{} < -22$) continuum-selected $z>6$ galaxies in the literature \citep{Cuby2003,Willott2013,Oesch2015}, indicating an average FWHM=450 km s$^{-1}$ for this population. These broad line widths undoubtedly assist in boosting the \Lya{} visibility of UV-bright galaxies at $z\gtrsim7$.
    
    \item A contributing factor to the broad \Lya{} profiles of massive, UV-bright $z>6$ galaxies may be that their \Lya{} emission is produced in gas spanning a wide range of motions, as evidenced by our ALMA data. In the two REBELS galaxies showing extremely broad \Lya{} profiles, we also observe very broad [CII] emission (FWHM=300--520 km s$^{-1}$) indicating that these are composite galaxies containing multiple gas clumps moving with large peculiar motions \citep[e.g.][]{Bowler2017,Carniani2018}. If these high-velocity gas clumps also produce \Lya{} emission, the intrinsic \Lya{} profile will mirror that of [CII] before being further broadened by resonant interactions with outflowing material. The large peculiar motions of these clumps might also produce \Lya{} profiles with multiple peaks redward of systemic velocity. Such profiles are different in nature from those showing a peak blueward of systemic which signal low HI column density channels on the near side of the galaxy \citep[e.g.][]{Gazagnes2020}. We suggest that systemic redshifts from [CII] and other non-resonant lines are critical to interpret the physical origin of multi-peaked \Lya{} lines.
    
    \item We find no strong trend between the strength of \Lya{} emission and the [CII] luminosity at fixed SFR in our REBELS sample. The three galaxies with relatively weak \Lya{} emission (EW$<$10 \AA{}) show an average [CII] luminosity to SFR ratio of (0.09$\pm$0.04)$\times$10$^8$ L$_{\odot}$/(M$_{\odot}$ yr$^{-1}$), very similar to the average ratio of (0.06$\pm$0.02)$\times$10$^8$ L$_{\odot}$/(M$_{\odot}$ yr$^{-1}$) among the three more moderate \Lya{} emitters (EW = 10--20 \AA{}). This lack of an apparent trend may be due to the limited dynamic range in \Lya{} EWs among our Lyman-break selected sample, though improved statistics are also required to test for a modest correlation in this EW regime.
  
    \item Recent studies have demonstrated that faint galaxies in the vicinity of ionized bubbles at $z\sim7$ do not show strong \Lya{} emission \citep{Castellano2018}. While this could imply relatively small bubbles around bright sources, it may also reflect a luminosity dependence on the \Lya{} profile properties discussed in this paper. Faint ($\Muv{} \sim -19$) $z\sim7$ galaxies are expected to typically exhibit much smaller velocity offsets ($\DelvLya{} \sim 100$ km s$^{-1}$; \citealt{Mason2018_IGMneutralFrac}) and narrower line widths (FWHM$\sim$200--250 km s$^{-1}$; \citealt{Verhamme2018}) relative to that seen among extremely luminous ($\Muv{} < -22$) systems. These predictions imply $\approx$2$\times$ lower IGM transmission among faint $z=7$ galaxies at fixed moderate bubble size of $R = 0.5$ physical Mpc, helping explain why their emission is more difficult to detect.
    
\end{enumerate}
  
\section*{Acknowledgements}

Observations reported here were obtained at the MMT Observatory, a joint facility of the University of Arizona and the Smithsonian Institution.
RE sincerely thanks the MMT queue observers Michael Calkins, Ryan Howie, ShiAnne Kattner, and Skyler Self for their assistance in collecting the Binospec data, as well as Ben Weiner for managing the queue.
This paper is based on data obtained with the ALMA Observatory, under the Large Program 2019.1.01634.L. ALMA is a partnership of ESO (representing its member states), NSF(USA) and NINS (Japan), together with NRC (Canada), MOST and ASIAA (Taiwan), and KASI (Republic of Korea), in cooperation with the Republic of Chile. The Joint ALMA Observatory is operated by ESO, AUI/NRAO and NAOJ. 
We acknowledge assistance from Allegro, the European ALMA Regional Center node in the Netherlands.
This work is based [in part] on observations made with the Spitzer Space Telescope, which was operated by the Jet Propulsion Laboratory, California Institute of Technology under a contract with NASA. 
Based on data products from observations made with ESO Telescopes at the La Silla Paranal Observatory under ESO programme ID 179.A-2005 and on data products produced by CALET and the Cambridge Astronomy Survey Unit on behalf of the UltraVISTA consortium.
This work is based in part on data obtained as part of the UKIRT Infrared Deep Sky Survey.

RE and DPS acknowledge funding from JWST/NIRCam contract to the University of Arizona, NAS5-02015.
DPS acknowledges support from the National Science Foundation through the grant AST-1410155.
RJB and MS acknowledge support from TOP grant TOP1.16.057. 
SS acknowledges support from the Nederlandse Onderzoekschool voor Astronomie (NOVA).
RS and RAAB acknowledge support from STFC Ernest Rutherford Fellowships [grant numbers ST/S004831/1 and ST/T003596/1]. 
HI acknowledges support from NAOJ ALMA Scientific Research Grant Code 2021-19A and the JSPS KAKENHI Grant Number JP19K23462.
PAO acknowledges support from the Swiss National Science Foundation through the SNSF Professorship grant 190079 `Galaxy Buildup at Cosmic Dawn'. 
MA acknowledges support from FONDECYT grant 1211951, ``CONICYT + PCI + INSTITUTO MAX PLANCK DE ASTRONOMIA MPG19003'' and ``CONICYT+PCI+REDES 190194''.
PD and AH acknowledge support from the European Research Council's starting grant ERC StG-717001 (``DELPHI''). 
PD acknowledges support from the NWO grant 016.VIDI.189.162 (``ODIN'') and the European Commission's and University of Groningen's CO-FUND Rosalind Franklin program.
AF and AP acknowledge support from the ERC Advanced Grant INTERSTELLAR H2020/740120. Any dissemination of results must indicate that it reflects only the author's view and that the Commission is not responsible for any use that may be made of the information it contains. Partial support from the Carl Friedrich von Siemens-Forschungspreis der Alexander von Humboldt-Stiftung Research Award is kindly acknowledged (AF). 
LG and RS acknowledge support from the Amaldi Research Center funded by the MIUR program ``Dipartimento di Eccellenza'' (CUP:B81I18001170001). 
TN acknowledges support from Australian Research Council Laureate Fellowship FL180100060.

This research made use of \textsc{astropy}, a community-developed core \textsc{python} package for Astronomy \citep{astropy:2013, astropy:2018}; \textsc{matplotlib} \citep{Hunter2007_matplotlib}; \textsc{numpy} \citep{van2011numpy}; and \textsc{scipy} \citep{jones_scipy_2001}.

\section*{Data Availability}
 
The near- and mid-infrared imaging data underlying this article are available through their respective data repositories. See \url{http://www.eso.org/rm/publicAccess#/dataReleases} for UltraVISTA and VIDEO data, and \url{https://sha.ipac.caltech.edu/applications/Spitzer/SHA/} for IRAC data. 
The raw ALMA data are available via the science archive (program 2019.1.01634.L) and the MMT/Binospec data will be shared upon reasonable request to the corresponding author.

%%%%%%%%%%%%%%%%%%%%%%%%%%%%%%%%%%%%%%%%%%%%%%%%%%



\bibliographystyle{mnras}
\bibliography{paper_ref} % if your bibtex file is called example.bib

%%%%%%%%%%%%%%%%%%%%%%%%%%%%%%%%%%%%%%%%%%%%%%%%%%

%%%%%%%%%%%%%%%%% APPENDICES %%%%%%%%%%%%%%%%%%%%%

\appendix
 
%%%%%%%%%%%%%%%%%%%%%%%%%%%%%%%%%%%%%%%%%%%%%%%%%%

% Don't change these lines
\bsp	% typesetting comment
\label{lastpage}
\end{document}